\documentclass[aps,a4paper]{revtex4}

\usepackage{epsfig}
\usepackage{graphicx}
\usepackage{amsmath,amssymb,color}
\usepackage[english]{babel}
\usepackage[colorlinks=true, allcolors=blue]{hyperref}
\usepackage{tabularx}

\parskip=\medskipamount

\newcommand{\eq}[1]{(\ref{#1})}
\newcommand{\fig}[1]{Fig. \ref{#1}}
\newcommand{\tab}[1]{Table ~\ref{#1}}

\newcommand{\be}{\begin{equation}}
\newcommand{\ee}{\end{equation}}

\newcommand{\barr}{\begin{array}}
\newcommand{\earr}{\end{array}}

\newcommand{\beqn}{\begin{eqnarray}}
\newcommand{\eeqn}{\end{eqnarray}}

\newcommand{\bs}{\begin{subequations}}
\newcommand{\es}{\end{subequations}}

\newcommand{\bw}{\begin{widetext}}
\newcommand{\ew}{\end{widetext}}

\newcommand{\ve}{\mathbf}

\begin{document}

\title{Nearest-neighbour directed random hyperbolic graphs}

\author{I.A.~Kasyanov$^{1}$, P. van der Hoorn$^2$, D. Krioukov$^3$, M.V.~Tamm$^{4}$}

\affiliation{$^1$ Independent researcher, Tbilisi, Georgia;\\
$^2$ Eindhoven University of Technology, Eindhoven, Netherlands; \\
$^3$ Northeastern University, Boston, MA, USA; \\
$^4$ ERA Chair for Cultural Data Analytics, School of Digital Technologies, Tallinn University, Tallinn, Estonia; \\}

\date{\today}

\begin{abstract}
Undirected hyperbolic graph models have been extensively used as models of scale-free small-world networks with high clustering coefficient. Here we presented a simple directed hyperbolic model, where nodes randomly distributed on a hyperbolic disk are connected to a fixed number $m$ of their nearest spatial neighbours. We introduce also a canonical version of this network (which we call ``network with varied connection radius''), where maximal length of outgoing bond is space-dependent and is determined by fixing the average out-degree to $m$. We study local bond length, in-degree and reciprocity in these networks as a function of spacial coordinates of the nodes, and show that the network has a distinct core-periphery structure. We show that for small densities of nodes the overall in-degree has a truncated power law distribution. We demonstrate that reciprocity of the network can be regulated by adjusting an additional temperature-like parameter without changing other global properties of the network. 
\end{abstract}


\maketitle

\section{Introduction}

Reference and recommendation networks are ubiquitous\cite{viral,recom}: encyclopedia articles and scientific papers refer to each other, people recommend each other books, films and music, online shops are full of “people who like this also like that” recommendations. These recommendations constitute directed links between objects organizing them into a directed network. The resulting networks are substantially asymmetric: the rules according to which a node becomes a source of recommendation are different from those according to which it gets recommended. In many cases the number of recommendations given {\it from} a node is either strictly or effectively bounded, while the number of recommendations {\it towards} a node is unlimited. Accordingly, out-degree and in-degree distributions in such networks are very different: out-degree is relatively narrowly distributed, while in-degree distribution typically has a wide, often power-law tail.

A particular example of this type of networks is a network of free associations in a language\cite{kiss,nelson,dedayne,valba}. Typically, it is constructed as follows\cite{nelson,dedayne}. Test subjects receive a set of words (stimuli) and they are to provide a first word which came to their mind as a response to each of the stimuli. The results are aggregated into a directed network of associations weighted according to the frequency with which associations appear in the dataset. The resulting networks have narrow out-degree distributions but wide in-degree distributions with power-law tails \cite{valba}. Another nice example of a directed network with asymmetry between narrow in-degree and power-law out-degree is the network of mathematical theorems studied in \cite{dedeo}. Yet another example where natural asymmetry between in- and out-degree distributions arises is the system of links in encyclopedia (see, e.g., \cite{konnect}). However, typically in this case both distributions have power-law tails with unequal exponents: it is less probable to have a very large out-degree than a very large in-degree. 

History of network science is full of examples of how essential it is to have benchmark models of random graphs, which are able to reproduce some of the properties of the experimentally observed networks (see textbook presentation in \cite{dorog,newman_book,barabasi_book,jackson,krapivsky_book,barthelemy_book}). Emergence of the giant cluster was understood by Flory\cite{flory}, Erdos and Renyi\cite{erdos} based on a minimalistic model. Watts-Strogatz model \cite{ws} is essential for understanding the emergence of the small world effect. Barabasi-Albert\cite{ba} and other preferential attachment models\cite{krapivsky_pref} shed light on the emergence of power law degree distributions. Studying configuration \cite{conf} and exponential graph models \cite{ParkNewman} is essential to separate the effects of various topological invariants (degree, motif distributions, etc) on the properties of networks. Hyperbolic network models, also known as random hyperbolic graphs\cite{krioukov1,krioukov2,krioukov3} explain how power-law distributions, high clustering and small world properties coexist with each other, which is often the case in real-world networks. These models, as well as somewhat similar Apollonian networks \cite{apol1,apol2} and their generalizations\cite{zhang1,zhang2,bianc,tks} are the first equilibrium models, which unify these three properties. However, up till now  hyperbolic random graph models have been confined to undirected networks (except for two very recent papers \cite{boguna_dir},\cite{kovacs}, see the discussion section).    

Here we develop and study a simple model of a directed network with asymmetric degree distribution: narrow distribution of the out-degree and wide distribution of the in-degree. Our model is a hyperbolic generalization of nearest-neighbour models\cite{eucl1,eucl2,eucl3,eucl4} (see also reviews \cite{balobas,walters}) studied extensively for the case of Euclidean metric spaces (interestingly, nearest-neighbour graphs in high-dimensional Euclidean spaces are an important intermediate step in the construction of popular dimension-reduction algorithms such as diffusion maps \cite{diff}, t-SNE \cite{tsne} and UMAP \cite{umap}). We consider a disk in a hyperbolic space, drop a large number of points onto it uniformly at random, and then connect each point by directed links to a fixed number of its nearest neighbours. We study the limit of large networks, and show that the in-degree distribution of such a network is a truncated power law, and by adjusting parameters the power law region can be made arbitrarily wide. We also discuss how it is possible to regulate the structural parameters of the network, most importantly, the reciprocity of the network (i.e., fraction of bidirectional links in it) and the exponent of the power law distribution. 

The presentation is organized as follows. In section II we define the model, discuss it qualitatively and formulate the main results. We also define an auxiliary conjugate model, which we call ``network with varied connection radius''(VCR). This model has properties similar to those of the nearest-neighbour model but is more tractable analytically. In section III we turn to a more quantitative approach and provide analytical and numerical calculations for the location-dependent bond length and in-degree, and show that the nearest-neighbor network has a peculiar core-periphery structure. We end section III with deriving the truncated power law behavior of the overall in-degree distribution. 

In section IV we recall the definition of network reciprocity and calculate it for the nearest-neighbor and VCR networks. We show how reciprocity can be regulated by introducing an additional temperature-like parameter. Finally, in section V we summarize our results and discuss their possible applications and generalizations, including the control of the exponent of the in-degree distribution.  

In what follows we assume some familiarity with the concept of hyperbolic spaces with constant negative curvature (see, e.g., \cite{cannon} for an extended introduction). However, all the concrete formulae needed to understand the result are provided in the text to make it self-contained.

\section {Definition of the model and qualitative discussion}

\subsection {The $m$-nearest-neighbour ($m$-NN) network model}

In this section we give the definition of the nearest-neighbour network model and qualitatively discuss the bond length and the degree distribution in this network.

We start with a general definition of an $m$-nearest-neighbour ($m$-NN) network. Consider a set $V$ of $N$ points $x_1,..., x_N$ in some normed vector space, so that distances $||x_i - x_j||$ are defined for all $i,j$. Call an $m$-NN network a graph $G$ consisting of vertices $V$ and $mN$ directed edges, connecting each vertex to its $m$ nearest neighbours, i.e., to $m$ nodes, the distance to which is smallest. Clearly, the result is a directed graph with out-degree distribution $P_{out}(k)= \delta_{k,m}$. The average in-degree is also $m$, but the in-degree distribution may be quite non-trivial. 

Consider first a simple stochastic setting: namely, let points be distributed uniformly and independently at random with a given density $\nu$ per unit volume in an infinite $d$-dimensional space of constant curvature. This problem has been studied extensively in the case of Euclidean (zero-curvature) space\cite{balobas,walters}. In this case, the average network properties are translationally invariant, and both the bond length distribution and the in-degree distribution are rather narrow. Indeed, although the exact distribution of in-degree for arbitrary $m,d$ is unknown (see \cite{taowu} where $P_{in}(k=0)$ is calculated for $m=1, d=2$), it is easy to show that the probability of both a large bond length $r$ and a large in-degree $k$ decays at least exponentially in both Euclidean and hyperbolic space.

\begin{figure}
\includegraphics[width=7cm]{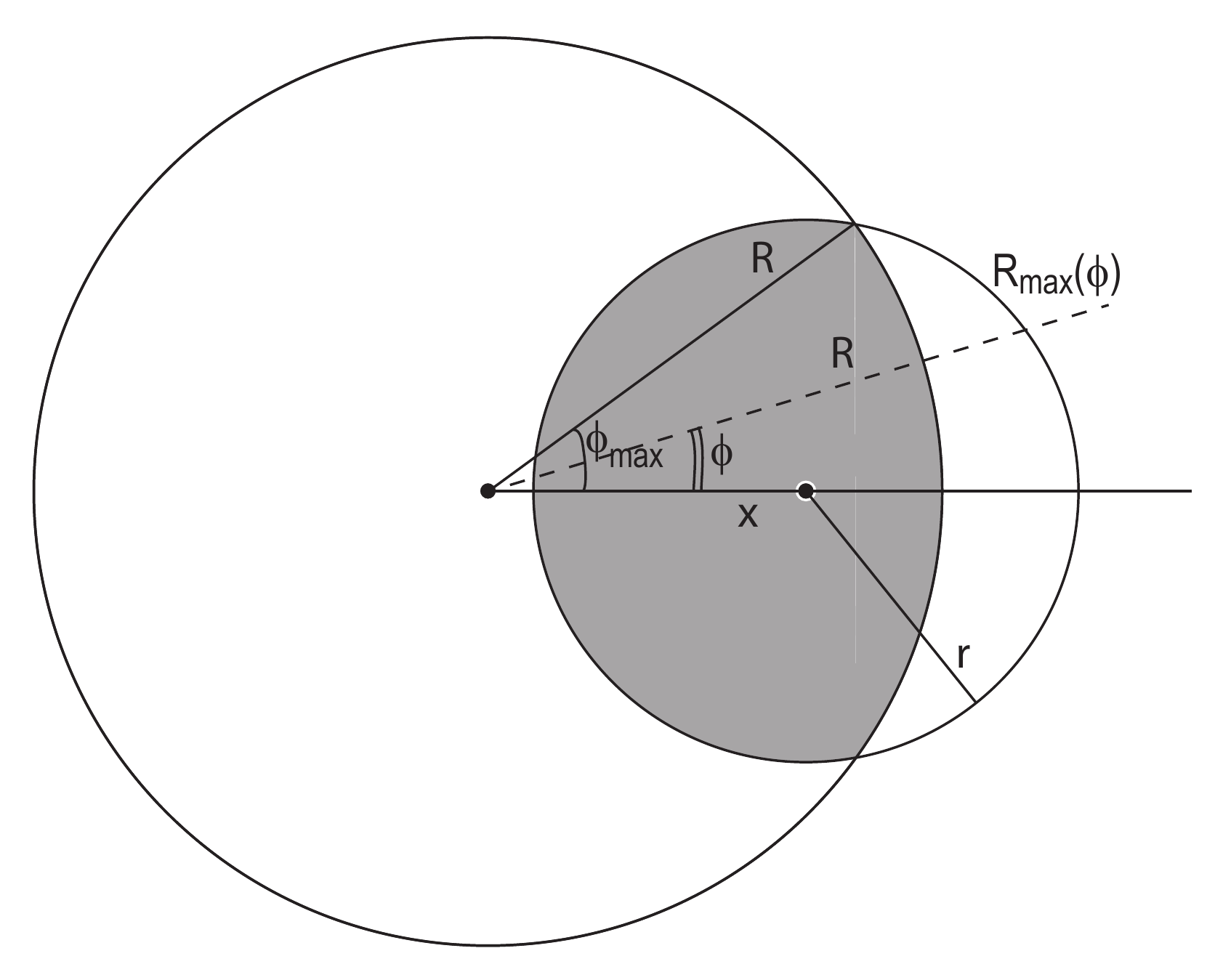}
\caption{Sketch of the overlap of two circles. If the circles intersect, the shaded ares $A(r,x,R)$ is smaller than the area of the small circle $A(r)$. The angle $\phi_{\max}$ and distance $R_{\max}(\phi)$ are also shown, see \eq{phimax}, \eq{Rmax}.}
\label{fig:sketch}
\end{figure}

Indeed, let $A(r)$ be the volume of a ball of radius $r$ (henceforth all distances are assumed to be hyperbolic unless mentioned otherwise; a \emph{ball} is then defined, as usual, as a set of points at distance no more than $r$ from the center, and a \emph{disk} is a ball in 2-dimensional space). Then, define distance $r_m$ as a solution of equation
\be
\nu A(r_m) = m,
\label{r_m}
\ee
i.e., within the ball of radius $r_m$ there are on average $m$ points. The radius $r_m$ has a meaning of ``typical distance'' to the $m$-th nearest neighbour or typical bond length. Indeed, the probability that there are less than $m$ points in the ball of radius $r$ (which is equal to the probability that the distance to the $m$-th neighbour is larger than $r$) is
\be
P_m (r) = \exp(-\nu A(r)) \sum_{k=0}^{m-1} \frac{(\nu A(r))^k}{k!},
\label{large_r}
\ee
Note now that in the expression
\begin{equation}
\exp(-\nu A(r)) = \sum_{k=0}^{\infty} \frac{(\nu A(r))^k}{k!}    
\end{equation}
the maximal term in the r.h.s. is located at $\nu A(r) = k$ and thus the distribution of the $m$-th nearest neighbor $p_m (r) = dP_m(r)/dr$ is localized around $r_m$. In what follows, we use $r_m$ as a qualitative estimate of the distance to $m$-th nearest neighbor, since the more conventional estimate, mean distance to $m$-th nearest neighbour
\begin{equation}
r_m^{\text{mean}} = \int_0^{\infty} r p_m(r) dr    
\end{equation}
is of the same order of magnitude but is a bit harder to calculate and depends on a particular form of $A(r)$ in a more complicated way. Note also that big deviations of the bond length from $r_m$ are extremely rare: 
$A(r) \sim r^{d}$ in $d$-dimensional Euclidean space, and $A(r) \sim \exp r$ for large $r$ in hyperbolic space, so in both cases $P_m(r)$ given by \eq{large_r} decays superexponentially, effectively limiting the possible values of $r$. This also means that in the translationally-invariant case the in-degree of a node in the $m$-NN model is limited. Indeed, the superexponential decay of \eq{large_r} means that there are essentially no bonds in the network longer than $ar_m$, where $a$ is some numerical constant of order 1. Then the in-degree is effectively limited by
\begin{equation}
    k_{in}^{\max} \sim \nu A(ar_m) =\left\{ 
    \begin{array}{ll}
    m a^{d-1}  &\text{ for Euclidean space;} \medskip \\
    m e^a  &\text{ for hyperbolic space.}
    \end{array}
    \right.
\end{equation}

The situation becomes more interesting for the case of a nearest-neighbour network in a bounded domain, where translation invariance breaks up and average properties of nodes are location-dependent.  Consider the simplest possible setting when points are distributed uniformly and independently at random with a given density $\nu$ inside a ball $O(R)$ of radius $R$ (here and in what follows whenever discussing hyperbolic space we, without loss of generality, set the space curvature $\zeta =1$, i.e. distance $R$ in the hyperbolic space is measured in the units of inverse space curvature and is therefore a dimensionless variable). In this case, the average length of a bond becomes dependent on the spatial position of its source, and the in-degree of a node depends on its radial coordinate (distance from the center of the ball). Indeed, consider the overlap of a ball of radius $r$ with center at radial coordinate $x$ and the ball $O(R)$. For $x+r>R$ the volume of this overlap $A(r,x,R)$ (grey area in the sketch \fig{fig:sketch}) is smaller than the volume of unrestricted ball $A(r)$ (the area of the small circle in \fig{fig:sketch}). Thus, the number of points inside $A(r_m,x,R)$ is on average smaller than $m$, and the typical distance to the $m$-th nearest neighbour becomes larger than $r_m$. Similarly to \eq{r_m} we define this typical distance $r_m(x)$ as a solution of equation
\be
\nu A(r_m(x), x, R) = m.
\label{r_mx}
\ee
For brevity, we preserve notation $r_m$ without explicit $x$ dependence for the limiting value of $r_m(x)$ in the bulk, so that $r_m = \lim_{R \to \infty} r_m(x)$ for any fixed $x$.

\begin{figure}
\includegraphics[width=15cm]{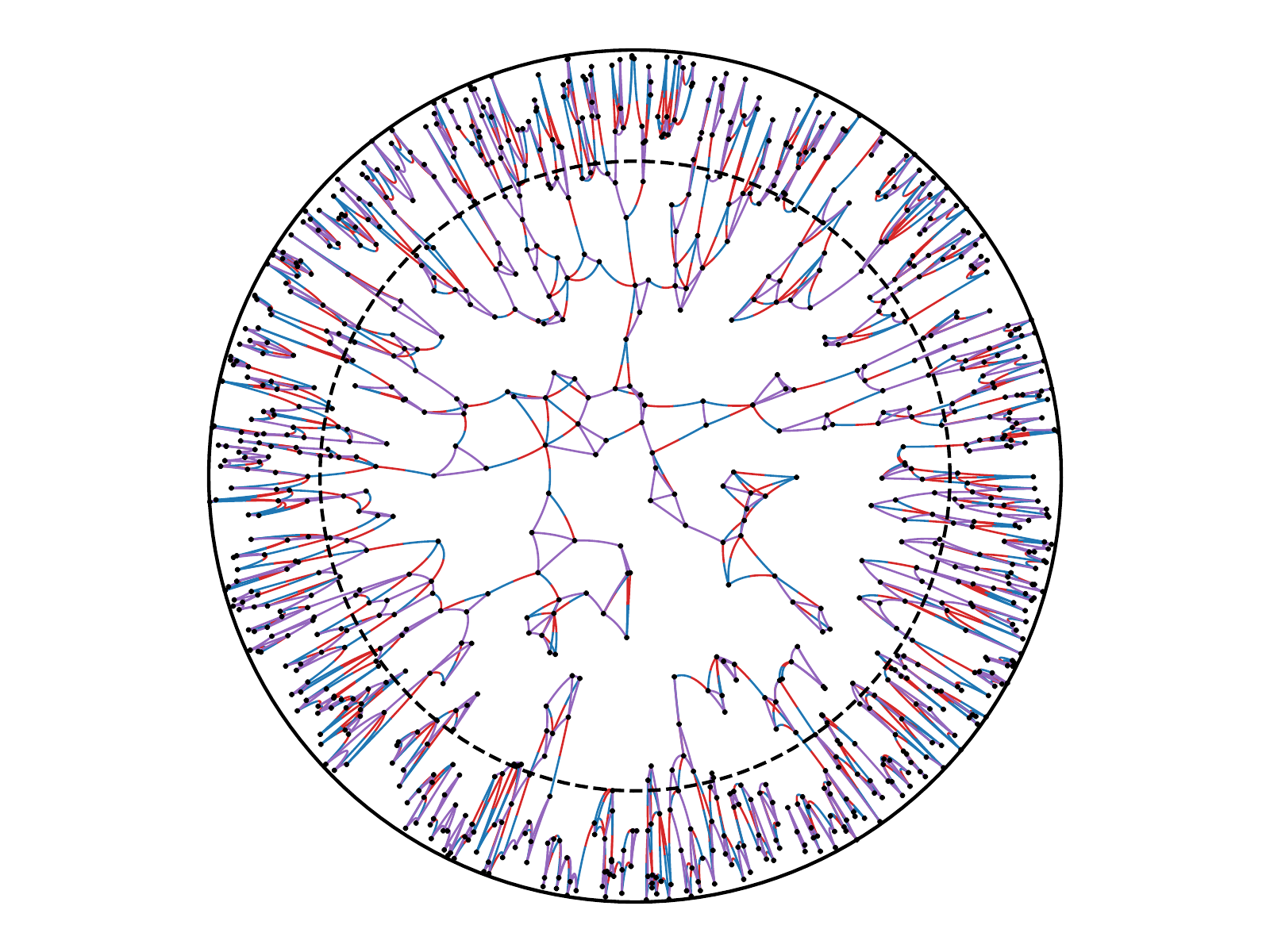}
\caption{Example of a directed nearest-neighbor network of $N=1000$ nodes on a hyperbolic disk. Each node is connected to $m=3$ nearest neighbors (in the colored version the source and target ends of each bond are shown with blue and red, respectively, bidirectional bonds are shown in purple). Note the core-periphery structure of the network: in the periphery the bonds are directed predominantly towards the center of the disk, in the core there is no predominant direction. The in-degree is maximal at the crossover from core to periphery, in the vicinity of the dashed black circle.}
\label{f:00}
\end{figure}

Consider now how this spatial dependence of the bond length influences in-degree. Points in the immediate vicinity of a boundary have on average an in-degree smaller then $m$ because part of the points, which would normally  connect to them in the infinite domain case are now located outside $O(R)$. Conversely, points at a distance somewhat larger than $r_m$ from the boundary have an increased average in-degree: apart from all the bonds of the length $r<r_m$ which they attain similarly to the points in the bulk, they also get a certain number of abnormally long incoming bonds from the points in the close vicinity of the boundary. Thus, in-degree depends on the spatial coordinate of the bond in a non-monotonic way: it equals $m$ when the distance from the boundary 
\be
\xi=R-x
\ee
is much larger than $r_m$, then increases to a larger number somewhere around $r_m$, then drops to a value smaller than $m$ in the immediate vicinity of the boundary. 

This boundary effect exists both in the Euclidean and the hyperbolic space. In the Euclidean space it is relatively small (numerically, in 2D we found a spatial variation of in-degree by a factor of roughly 2), and, even more importantly, in the {\it thermodynamic limit} $R\to \infty , \nu = const$ it affects an infinitesimal fraction of nodes. Indeed, the fraction of volume occupied by the boundary region tends to zero for large domains in the Euclidean space. The situation is dramatically different in the hyperbolic space  (see \fig{f:00})  where volumes of the ball $A(r)$ and of its surface $dA(r)/dr$ increase exponentially with $r$. In this case the fraction of volume taken by the boundary converges to a finite number of order
\be
1- \frac{\exp(R-r_m)} {\exp R} = 1 - \exp(-r_m).
\ee  
In turn, $r_m$ is a function of dimensionless density $\nu$, and becomes arbitrary large for small $\nu$ and arbitrary small for large $\nu$ (see \eq{r_m}). Thus, by changing the density of points one can cross over from the regime of bulk dominance for large $\nu$ to the regime of boundary dominance for small $\nu$. As we show below, in the boundary region the in-degree distribution depends exponentially on $\xi$ (compare to \cite{krioukov2}), which, combined with the exponential growth of the number of nodes with a distance from the origin, leads, in full analogy with \cite{krioukov2} to a power-law distribution of the in-degree 
\be
P_{in} (k) \sim k^{-3}
\label{-3}
\ee
within the boundary region. Since the boundary region has a finite width, this power-law remains truncated even in the thermodynamic limit with $k_{in}^{\max} \sim m/\nu$. 

\subsection {Network with varied connection radius (VCR)}

Before proceeding further it is instructive to introduce an auxiliary network model, which plays the role of a canonical counterpart of the $m$-NN network model.  Consider once again a set of points distributed uniformly and independently at random with a given density $\nu$ inside $O(R)$, and for each point with radial coordinate $x$ add outgoing bonds to all points at a distance no more then $r_m(x)$ from it, with $r_m(x)$ given by equation \eq{r_mx}. Note that this model bears some similarity with \cite{gracar}, although here the rules governing connection of nodes are inhomogeneous in space, not in time. Whereas the out-degree of the $m$-NN network is $P_{out}(k)=\delta_{k,m}$, this auxiliary ``network with varying connection radius'' has a Poisson out-degree distribution with coordinate-independent mean equal to $m$:
\begin{equation}
    P_{out}^{\text{MF}}(k) = e^{-m} m^k/k!
\end{equation}
As we show below, the average in-degree of a node in this network is relatively easy to calculate: it is determined simply by the geometry of the domain and does not need averaging over simultaneous positions of several randomly located particles. On the other hand, the network with varied connection radius and the $m$-NN network are dual in a way similar to duality between canonical and microcanonical ensembles. Indeed, in $m$-NN network the out-degree is strictly fixed, while the distance to $m$-th neighbour is fluctuating around its (position-dependent) average $r_m(x)$ while in the network with varied connection radius the situation is inverted:  the maximal length of the bond is fixed and the total number of outgoing bond is fluctuating. For large $m$ the relative fluctuations are expected to be small and models are expected to converge in complete analogy to the canonical and microcanonical ensembles giving the same result in thermodynamic limit. 

In what follows we consider these two models in parallel, leveraging the relative analytical simplicity of the model with varied connection radius. In the next section we provide analytical and numerical calculations supporting and expanding the qualitative analysis of the two models presented above. 

\section{Quantitative analysis of the nearest-neighbour model}

\subsection{Circles overlap $A(r,x,R)$}

We start with finding the explicit from of the function $A(r,x,R)$ and based on that study the asymptotic behavior of the solution of \eq{r_mx}.

In what follows we use polar coordinates centered in the center of $O(R)$.  Recall that the infinitesimal area element $dA$ in hyperbolic polar coordinates is
\be
dA = \sinh \rho d\rho d\theta,
\ee
and thus circumference and area of a disk of radius $r$ are
\be
\Pi (r) = dA(r)/dr = 2\pi \sinh r;\;\;\; A(r) = 2 \pi (\cosh r-1),
\label{circle}
\ee
respectively (recall that all the lengths are measured in the units of inverse space curvature and are therefore dimensionless). This allows to solve equation \eq{r_m} explicitly, getting the following expression for the typical length of the bond to $m$-th nearest neighbour in the bulk
\be
r_m = \cosh^{-1}\left[\frac{m}{2\pi \nu} +1 \right] \approx \ln \left[\frac{m}{\pi \nu} +2 \right],
\label{rm_explicit}
\ee
where the last equality is valid in the small density ($m/\nu \gg 1$) limit.

Recall also the hyperbolic cosine theorem
\be
    \label{cosine}
    \cosh a = \cosh b \cosh c - \sinh b \sinh c \cos \alpha,
\ee
which connects the length of the sides of a triangle $a,b,c$ and the angle $\alpha$ opposite to the side $a$.

\begin{figure}
\includegraphics[width=17cm]{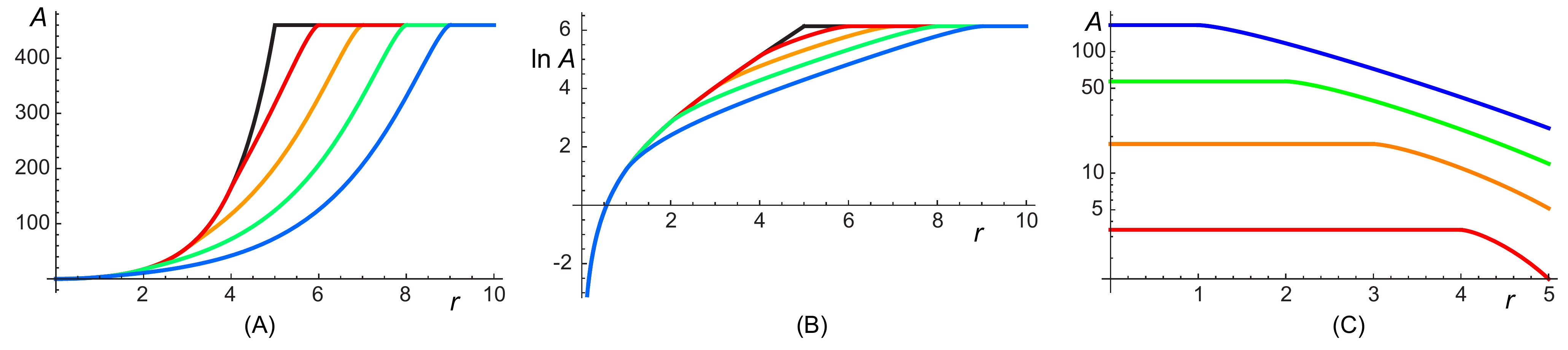}
\caption{The area of the overlap $A(r,x,R)$ of two circles as a function of the radius of the inner circle $r$ for $R=5$ and various coordinates of the center of smaller circle $x$: (from top to bottom) $x=$ 0(black), 1(red), 2 (orange), 3(green), 4(blue, bottom) plotted in linear (A) and logarithmic (B) scale. (C) The area of the overlap $A(r,x,R)$ as a function of the position of the center of the smaller circle $x$ for $R=5$ and (from bottom to top) $r=$ 1 (red), 2 (orange), 3 (green) and 4 (blue).}
\label{fig:area}
\end{figure}

These formulae are enough to explicitly calculate $A(r,x,R)$, the overlap of a disk with radius $r$ and center at $x$ and the underlying disk $O(R)$. Clearly, there are two simple limiting cases. If $r \leq R-x$,  the small circle of radius $r$ is completely inside the disk, and thus
\be
A(r,x,R) = 2\pi (\cosh r -1).
\label{asmall}
\ee   
If $r \geq R+x$, in turn, the whole $O(R)$ is inside the circle of radius $r$ and  
\be
A(r,x,R) = 2\pi (\cosh R -1).
\label{alarge}
\ee   
Meanwhile, for $R-x < r< R+x$ the two disks intersect and the area can be calculated as (see \fig{fig:sketch})
\be
        A(r,x,R) = 2\pi(\cosh r - 1) -
        2\int\limits_0^{\phi_{\max}}d\phi\int\limits_R^{R_{\max}(\phi)} \sinh z dz  = 
        2\pi(\cosh r - 1) +2\phi_{\max}\cosh R  -
        2\int\limits_0^{\phi_{\max}}\cosh R_{\max}(\phi) d\phi,
\label{over1}
\ee
where the angle $\phi_{\max}$, the angular coordinate of the intersection of the circles, can be calculated from the hyperbolic cosine theorem
\be
        \phi_{\max} = \arccos\left(\dfrac{\cosh x \cosh R - \cosh r}{\sinh x \sinh R}\right),
\label{phimax}
\ee
as well as the radial coordinate of the point at the outer part $R_{\max} (\phi)$ of the smaller circle for any given $\phi$ 
\be
R_{\max} (\phi) = \text{arccosh} \left[ \dfrac{\cosh x \cosh r + \sinh x \cos \phi \sqrt{\sinh^2 r - \sin^2 \phi \sinh^2 x}}{1 +  \sin^2 \phi \sinh^2 x}\right]
\label{Rmax}
\ee
The integral in \eq{over1} can be calculated explicitly:
\be
\begin{array}{rll}
\displaystyle \int\limits_0^{\phi_{\max}}\cosh R_{\max}(\phi) d\phi &=& \displaystyle \cosh r \left[\frac{\pi (1-\text{sign}(\tan \phi_{\max}))}{2} + \arctan\left(\cosh x \tan \phi_{\max}\right)\right] + \medskip \\
& + & \displaystyle \int_0^{\sinh x \sin \phi_{\max}} \frac{\sqrt{\sinh^2 r - u^2}}{1+u^2} du = \medskip \\
& = &\displaystyle  \cosh r  \left[\frac{\pi (1-\text{sign}(\tan \phi_{\max}))}{2} + \arctan\left(\cosh x \tan \phi_{\max}\right)\right] - \medskip \\
& - & \displaystyle \arctan \left[ \frac{\sinh x \sinh R}{\cosh r \cosh R - \cosh x} \sin \phi_{\max} \right] + \medskip \\
& + & \displaystyle \cosh r \arctan \left[ \cosh r  \frac{\sinh x \sinh R}{\cosh r \cosh R - \cosh x} \sin \phi_{\max} \right]
\end{array}
\label{over_integral}
\ee
Equations \eq{asmall}-\eq{over_integral} collectively define the overlap area. \fig{fig:area} shows the $A(r,x,R)$ as a function of $r$ and $x$ for given $R=5$. Note that in the overlap regime $R-x<r<R+x$ the dependence of $A$ on both $x$ and $r$ is roughly exponential (at least if radius $r$ is not too small). Since this regime is the most relevant one for the dependence of the length of the bond on location, let us discuss it in more detail.

\subsection{Asymptotic of the bond length}

\begin{figure}
\includegraphics[width=17cm]{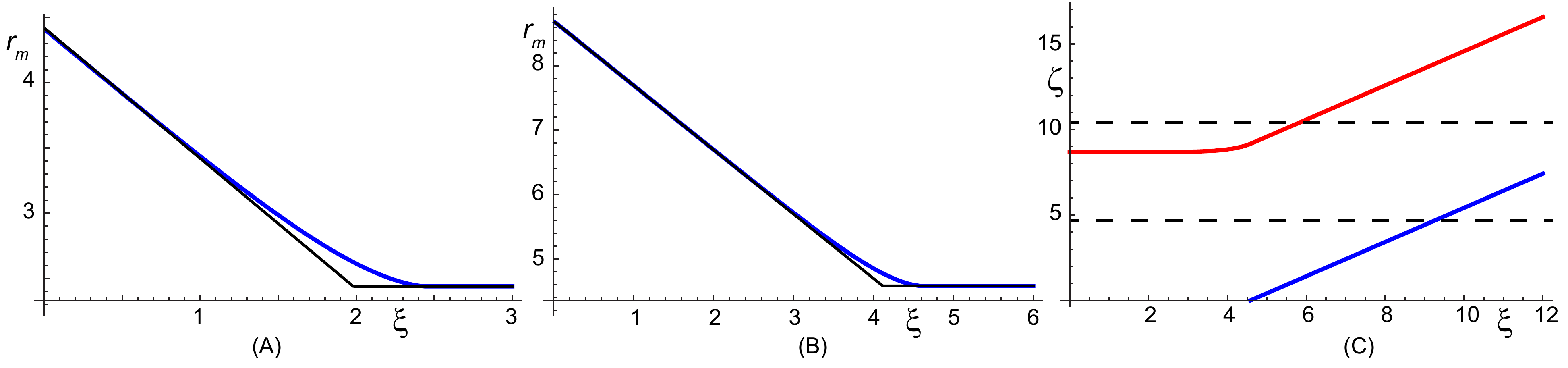}
\caption{The length $r_m$ of a typical bond to $m$-th nearest neighbour as a function of the distance $\xi$ from the source of the bond to the boundary of the disk, $m=3, \nu = 0.1$ (A) and 0.01 (B). Thick blue line corresponds to the solution of equation $\nu A(r_m,\xi) = m$ in the limit of large $R$ (i.e., for $A(r,m)$ given by \eq{Approx_A}); thin black line is the approximate solution given by \eq{rmx_approximate}. (C) the left hand side of equations \eq{ximinmax} for $\xi_{\min}$ (red, top) and $\xi_{\max}$ (blue, bottom) for $m=3, \nu = 0.01$. In the periphery, for $\zeta<r_m(0)$ (lower dashed line), $\xi_{\min}=0$, in the core, for $\zeta>2r_m(\infty)$ (upper dashed line), $\xi_{\min}=\zeta-r_m(\infty)$. There is a narrow intermediate region between these two regimes.
}
\label{fig:rmxi}
\end{figure}

It is instructive to study asymptotic behavior of the expressions \eq{over1}, \eq{phimax}, \eq{over_integral} in several steps. First, consider the thermodynamic limit $R\to \infty$. In this case $\phi_{\max}$  becomes exponentially small in $R$:   
\be
   \phi_{\max} \approx \tan \phi_{\max} \approx \sin \phi_{\max} \approx  2\sqrt{2} \sqrt{\cosh r - \cosh \xi}\, e^{\xi/2} e^{-R}  ,  \;\;\; \xi = R-x,
\ee
which allows to rewrite the expression for the area $A(r,\xi)$ in a simplified form
\be
A(r,\xi) = \left\{ 
\begin{array}{ll} 2\pi(\cosh r - 1) & \text{    for   } r\leq \xi, \medskip \\
2\pi(\cosh r - 1) +2\sqrt{2 (\cosh r - \cosh \xi )\, e^{\xi}}  - 2 \cosh r \arctan \left[ \sqrt{2 (\cosh r - \cosh \xi ) \, e^{-\xi}}\right] + &  \medskip \\
 \displaystyle  +\; 2 \arctan \left[ \frac{\sqrt{2 (\cosh r - \cosh \xi ) \, e^{-\xi}}}{\cosh r - e^{-\xi}}\right] - 2 \cosh r \arctan \left[\cosh r \frac{\sqrt{2 (\cosh r - \cosh \xi ) \, e^{-\xi}}}{\cosh r - e^{-\xi}}\right]+ & \text{    for   } r> \xi. \medskip \\
  \displaystyle  +\; O(e^{-R}) &
 \end{array}\right.
\label{Approx_A}
\ee
Introduce now $u=r-\xi $.  If $u$ is positive and $\xi$ is sufficiently large (i.e., if the circle does intersect with the boundary and its center is not too close to it), $\cosh r \approx e^{\xi +u}/2 \gg e^{-\xi}$, which leads to a further simplification
\be
A(r,\xi) \approx e^{\xi} \left(2 \sqrt{e^u - 1} +e^{u}\left[ \pi   - 2 \arctan \sqrt{ e^u-1}\,  \right]\right) - 2\pi  \approx 4 e^{\xi+u/2} -2\pi,
\label{Approx_A2}
\ee
where the last equality is valid in the $e^u \gg 1$ limit. Thus, we arrive at the following approximate solution of \eq{r_mx} giving the typical distance to the $m$-th nearest neigbor from the point at distance $\xi$ from the boundary of the disk
\be
r_m (\xi) = 2 \ln\left[\frac{m}{4\nu} +\frac{\pi}{2}\right] -\xi = 2r_m - \xi +2\ln(\pi/4),
\label{rmx_approximate}
\ee
where $r_m$ is the limiting value of $r_m (\xi)$ at a large distance from the boundary and is given by \eq{rm_explicit}.
This very simple result is valid up to exponentially small corrections if all relevant exponents are simultaneously much larger than 1:
\be
e^R \gg 1,\,\,  e^\xi \gg 1,\,\, e^{r_m(\xi) -\xi} \gg 1, \,\, e^{r_m(\xi) -r_m} \gg 1,
\ee
that is to say, if the disk is large, the density of points is small, and the position of point in question is not too close to the boundary and not too close to the line separating the core and periphery regions. Importantly, the solution is linear in $\xi$. \fig{fig:rmxi} shows the numerical solution of \eq{r_mx} for $A$ given by \eq{Approx_A} and the approximation \eq{rmx_approximate}. It is seen that while for $m/\nu \sim 1$ the approximation \eq{rmx_approximate} is off, it becomes better and better for smaller densities, and for $m/\nu \gtrsim 10^2$ it works very well for most of the region $0<\xi<r_m+2\ln (\pi/4)$ except for small deviations when $\xi$ approaches the upper limit. These deviations remain finite as $\nu$ approaches zero and $r_m$ diverges.

Summing up, the typical length of the bond to $m$-th nearest neighbour depends on the position of the source in the following way: (i) if the source is more than $r_m$ away from the boundary, it is constant and equal to $r_m$; (ii) when the source point approaches the boundary, the length of the bond grows linearly with the distance $\xi$ to the boundary, according to \eq{rmx_approximate}; (iii) there is a narrow region of finite width in the vicinity of $\xi = r_m$, which smoothly connects these two regimes. 

In what follows we discuss the implications of this varied bond length on the {\it in-degree} of the nearest-neighbour network.

\subsection{In-degree in the network with the varied connection radius}

In the previous section we discussed how the typical length of the bond depends on the position of its {\it source} point $\ve x =(x,\phi)$. In this section we are going to study the properties of the bonds with a given position of the { \it target} point $\ve y=(y,\psi)$. It is easiest to consider first the auxiliary ``model with varied connection radius'', i.e. the model where a bond from $\ve x$ to $\ve y$ exists if and only if 
\be
r = ||\ve x - \ve y|| \leq r_m(x),
\ee
where $r_m(x)$ is given by \eq{r_mx}. Without loss of generality it is convenient to set $\psi = 0$, and to introduce also the distance from the target point to the boundary of the disk 
\be
\zeta = R-y.
\ee
Given that the points are distributed independently at random with a given density $\nu$, average in-degree of a point located at $\ve y$ equals
\be
\bar{k}_{in} (y) = \nu B_m (y) ,
\label{kin}
\ee
where $B_m (y)$ is the area of the domain ${\cal B}_m(y)$ defined as
\be
\ve x \in {\cal B}_m(y) \text{ if and only if } ||\ve x - \ve y|| \leq r_m(||\ve x||) \text{ and }  \ve x \in O(R), 
\ee
and for each given $y$ the in-degree has a Poisson distribution with average  $\bar{k}_{in} (y)$.

In the polar coordinates associated with the center of the disk the area $B_m (y)$  can be written formally as follows
\be
B_m (y) =  \int\limits_{x_{\min}}^{x_{\max}} \int\limits_{-\phi_{\max}}^{\phi_{\max}} \sinh x d\phi d \xi = 2 \int\limits_{x_{\min}}^{x_{\max}} \phi_{\max} (x,y) \sinh x  d \xi  
\label{barea}
\ee
where $\phi_{\max}$ is the value of the angular coordinate $\phi$ for which the distance between points $(x,\phi)$ and $(y,0)$ equals $r_m(x)$, while $x_{\min}$ and $x_{\max}$ define the range of coordinates of the source points from which the target at $y$ is accessible (provided the angular coordinates match). They are the solutions of equations
\be
y -x_{\min} = r_m(x_{\min});\;\;\; x_{\max}-y = r_m(x_{\max}).  
\label{xminmax}
\ee
 In terms of distances $\xi =R-x, \zeta = R-y$ from the boundary of the disk the equations \eq{xminmax} can be rewritten as
\be
\zeta = \xi_{\min} + r_m(\xi_{\min}); \;\; \zeta =  \xi_{\max} - r_m(\xi_{\max}) 
\label{ximinmax}
\ee
\fig{fig:rmxi} (C) shows the behavior of the left hand side of \eq{ximinmax}. Clearly, with respect to $\zeta$ there are two different regimes, corresponding to the core and the periphery of the network. 

In the core part, when $\zeta \geq 2r_m$, the accessible region is $x \in (y - r_m, y+r_m)$. In this case, the attachment radius is constant and equal to $r_m$ throughout the whole accessible domain, so ${\cal B}_m(y)$  is simply a circle of radius $r_m$,  so that one immediately gets  from the definition of $r_m$ \eq{r_m} that $\bar{k}_{in} (y) = m$. 

\begin{figure}
\includegraphics[width=17cm]{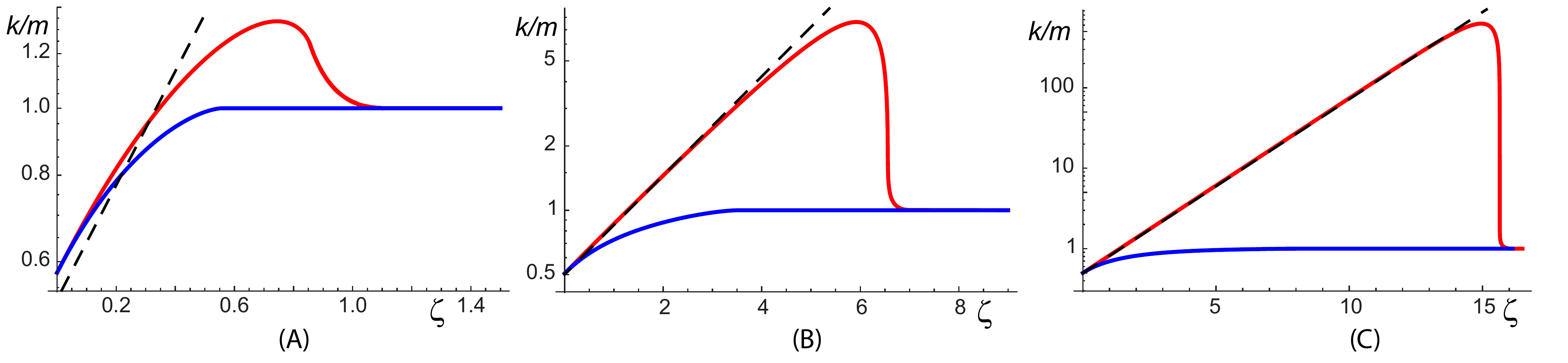}
\caption{The reduced average in-degree $\bar{k}_{in} (\zeta)/m$ (\eq{kin}, red, top) and bi-directional degree $\bar{k}_{\rightleftarrows} (\zeta)/m$ (\eq{bidegree}, blue, bottom) of the  network with varying connection radius as a function of the distance from the boundary $\zeta$ for (a) $m/\nu = 1$, (b) $m/\nu = 10^{2}$,  and (c) $m/\nu=10^{4}$. Dashed line indicates the approximate in-degree in the periphery region given by \eqref{mf_indegree_est}.}
\label{mf_indegree}
\end{figure}

In the periphery, when  $\zeta < 2r_m$ the  maximal length of the incoming bonds is source-dependent, so ${\cal B}_m(y)$ has a non-trivial shape, and its area is to be calculated using \eq{barea}. The angle $\phi_{\max}$ is obtained from the hyperbolic cosine theorem \eq{cosine}
\be
\cos \phi_{\max} = \frac { \cosh x \cosh y-\cosh r_m(x)}{ \sinh x \sinh y}; \;\; \psi_{\max} = 2\sqrt{2} e^{-R}\sqrt{e^{\xi + \zeta} (\cosh r_m(\xi)-\cosh (\xi -\zeta))} + O(e^{-2R}),
\ee
where we switched from $x,y$ to $\xi, \zeta$ notation, and used the expansion in powers of $e^{-R}$. Thus, in the leading order in $e^{-R}$ one gets 
\be
B_m (\zeta) = 2 \sqrt{2} \int\limits_{\xi_{\min}}^{\xi_{\max}} \sqrt{e^{ \zeta-\xi} (\cosh r_m(\xi)-\cosh (\xi -\zeta))} d \xi  + O(e^{-R})
\label{barea2}
\ee
where $r_m(\xi)$ is a solution of $A(r_m, \xi) = m/\nu$, $A(r_m, \xi)$ is given by \eq{Approx_A}, and the limits of integration are 
\be
\begin{array}{rll}
\xi_{\max} &= & \zeta + r_m, \medskip \\
\xi_{\min} & = & \left\{  
\begin{array}{ll}
0 & \text{  for }  \zeta \leq r_m(0), \medskip \\
\text{solution of  }  A(\zeta -\xi_{\min},\xi_{\min}) = m/\nu \;\;& \text{  for }  r_m(0)<\zeta < 2 r_m.
\end{array}
\right.
\end{array}
\ee 
This expression is easy to calculate numerically, the results are shown in \fig{mf_indegree}. In the core in-degree equals $m$, while in the periphery it grows exponentially with the distance from the boundary, spanning 3 orders of magnitude for $m/\nu = 10^4$, in between, there is a narrow crossover region where in-degree drops drastically with increasing $\zeta$. 

The exponential asymptotic in the periphery can easily be extracted by substituting approximation \eq{rmx_approximate} for $r_m(\xi)$. One gets
\be
B_m (\zeta) \approx 2 \int\limits_{0}^{\xi_max} \sqrt{e^{ \zeta-\xi} \left(\frac{\pi^2}{16}e^{2r_m-\xi}-\cosh(\xi -\zeta)\right)} d \xi  \approx 2 \int\limits_{0}^{\infty} \frac{\pi}{4}e^{r_m-\xi+\zeta/2} d\xi = \frac{\pi}{2} e^{r_m+\zeta/2} \approx \frac{1}{2} \frac{m}{\nu} e^{\zeta/2},
\label{mf_indegree_est}
\ee
where we used the fact that the integral is controlled by its lower bound, that $e^{2r_m-\xi}\gg\cosh(\xi -\zeta)$ in its vicinity, and that $m/\nu = 2\pi (\cosh r_m -1) \approx \pi \exp r_m$. This asymptotic is shown with a dotted line in \fig{mf_indegree}. 

The in-degree of the points located exactly at the boundary is $m/2$. The maximal in-degree of a network is reached at $\zeta_m = 2r_m - a$ where $a$ numerically is close to 1. Thus, the maximal in-degree of the network scales in the leading order as
\be
k_{in}^{\max} \approx m e^{\zeta_m/2} = C \times m e^{r_m} = C_2 \times \frac{m^2}{\nu}
\label{kinmax}
\ee
with the constant $C_2 \approx (2\pi)^{-1} \exp(-a) \approx 0.06$.

\subsection{In-degree distribution of the $m$-NN network}

The network with varied connection radius described above has an advantage of being relatively easy to treat analytically. The analysis of the in-degree in the $m$-NN network is a bit more cumbersome. However, the result is qualitatively very similar.

The average in-degree of a point located at $\ve y = (R-\zeta, 0)$ in the $m$-NN network can be formally written down as follows. For each pair of nodes define $m_{ij}$ - the position of node $i$ in the ordered   list of neighbours of the node $j$, that is $m_{ij} = 1$ if $i$ is the nearest neighbour of $j$, $m_{ij}=2$ if it is second nearest neighbour of $j$, etc. Then the in-degree of node $i$
\be
k_{in}^{(i)} = \sum_{j\neq i} \mathbb{I}(m_{ij} \leq m),
\label{kin_mnn}
\ee
where $\mathbb{I}(x)$ is the indicator function. Now, fix the position of $i$-th node at $\ve y$ and take the average of in-degree. One gets
\be
\bar{k}_{in} (\zeta) = \sum_{j\neq i} \sum_{k\leq m} \Pi(k,j|\zeta) = \int_{O(R)} \nu \sum_{k\leq m} \pi_{k} (\ve x|\zeta) d\ve x
\label{k_circle_0}
\ee
where $\Pi(k,j|\zeta)$ is the probability that the node at $\ve y = (R-\zeta, 0)$ is exactly $k$-th nearest neighbour of the $j$-th node, $\pi_{k} (\ve x|\zeta)$ is the similar probability for a node located at $\ve x$ and we took into account that nodes are distributed uniformly at random within $O(R)$ with density $\nu d \ve x$. The probability $\pi_k$ is the same as the probability that there are exactly $k-1$ points within the area $A(||\ve x - \ve y||,x,R)$ and is thus given by the Poisson distribution
\be
\pi_k(y,\ve x,\nu,R) = \frac{\left[\nu A(||\ve x - \ve y||,x,R)\right]^{k-1}}{(k-1)!} \exp (-\nu A(||\ve x - \ve y||,x,R)),
\label{pi_circle}
\ee
where the function $A(r,x,R)$ is defined by \eq{asmall}-\eq{over_integral}, and \eq{k_circle_0} can be rewritten as:
\be
\bar{k}_{in}(y,m, \nu,R) = \int_0^R \int_{-\pi}^{\pi}  \nu \sinh x dx d\psi \sum_{k\leq m} \pi_k(y, \ve x,R,\nu) = \int_0^R \int_{-\pi}^{\pi}  \nu \sinh x dx d\psi \frac{\Gamma(m,\nu A(||\ve x - \ve y||,x,R))}{m!},
\label{k_circle}
\ee
where $\Gamma(k,x)$ is the upper incomplete gamma-function. 

For large $R$ the integrand in the right hand side is localized in the vicinity of the position of the target node. This allows to rewrite in the limit $R \to \infty$ 
\be
\bar{k}_{in}(y,m, \nu) = \lim_{R\to\infty} \bar{k}_{in}(y,m, \nu,R) = \int_{0}^{\infty} e^{-\xi} d\xi \int_{-\infty}^{\infty} d \omega  \frac{\nu \Gamma(m,\nu A(||\ve x - \ve y||,\xi))}{m!}.
\label{k_circle_lim}
\ee
where we introduced linear coordinate along the boundary, $\omega = \psi \sinh R$, and the function $A(r,\xi)$ is given by \eq{Approx_A}.

\begin{figure}
\includegraphics[width=10cm]{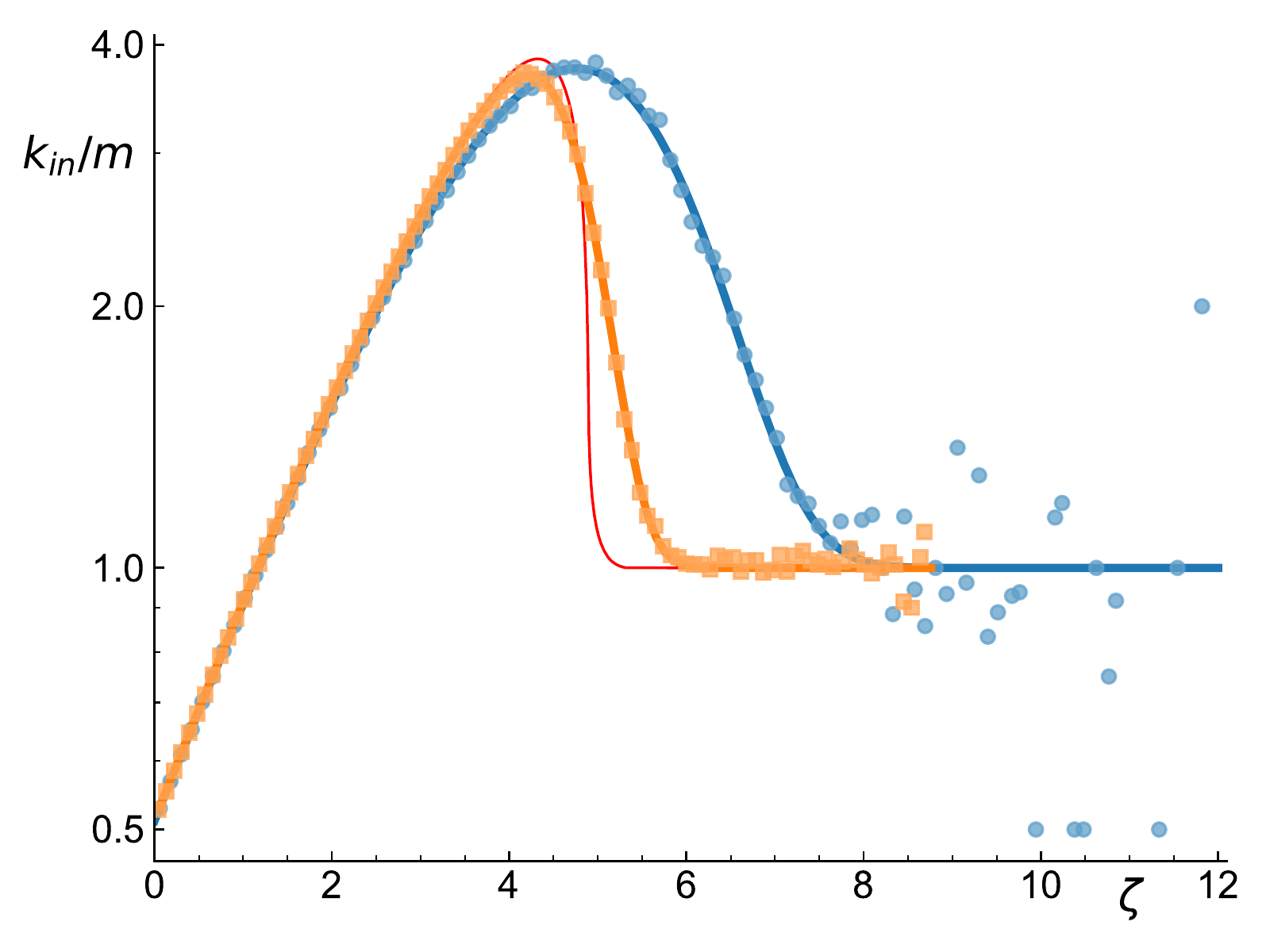}
\caption{Average in-degree of a node in the $m$-NN network as a function of the distance to the boundary $\zeta$ for $m/\nu = 40$. Thin red line corresponds to the result \eq{barea2} for the VCR network, thick blue (right) and orange (left) lines - to the $m$-NN networks with $m=2, \nu =0.05$ and with $m=20, \nu=0.5$, respectively. The lines are obtained by numerical evaluation of \eq{k_circle}, the blue circles and orange squares are the results of numerical simulations averaged over 100 networks of $10^4$ nodes (significant fluctuations for large $\xi$ are due to the fact that fraction of nodes with large $\xi$ decays as $e^{-\xi}$). }
\label{indegree_varm}
\end{figure}

\fig{indegree_varm} shows the behavior of \eq{k_circle_lim} for varying $m$ and fixed $m/\nu$, as well as the results of corresponding numerical simulations (see Appendix for the details of the simulation procedure). Qualitatively the behavior of the in-degree of $m$-NN networks given by \eq{k_circle_lim} is, indeed, similar to that of the VCR network given by \eq{barea2}: there are still the core region with $\bar{k}_{in} = m$ and the periphery region with exponential dependence of the in-degree on the distance to the boundary $\bar{k}_{in}(\zeta) = (m/2) \exp(\zeta/2)$. The only difference is in the width of the crossover region, connecting these two asymptotic behaviors: it is wider for smaller $m$ due to the fluctuations in the bond length. In the $m/\nu\to \infty$ limit the behavior of the $m$-NN network converges, as expected, to that of the network with varying connection radius, provided that the ratio $m/\nu$ is the same. 

\fig{indegree_varnu} shows how the behavior of the in-degree changes with changing $\nu$ and fixed $m$. Similarly to the VCR networks, the width of the periphery region and the maximal in-degree of the network increase with decreasing $\nu$, the maximal in-degree scaling according to \eq{kinmax}.

\subsection{Global in-degree distribution of the $m$-NN network}

In order to understand the global in-degree distribution of the $m$-NN network (i.e., summed over all nodes with various spatial positions) let us first discuss how the nodes of the network are distributed between the core and periphery regions. The fraction of nodes with a given distance $\zeta$ to the boundary is 
\be
f(\zeta) d\zeta = \frac{\sinh(R-\zeta)}{\sinh R} d\zeta = e^{-\zeta}d\zeta + O(e^{-R}) 
\label{fzeta}
\ee
Thus, in the limit $R\to\infty$ the fraction of the nodes in the core of the network is
\be
f_{core} = \int_{2r_m}^{\infty} f(\zeta) d\zeta = e^{-2r_m} =  \left( y +1 - \sqrt{y(y+2)}\right)^2 ,\;\;y=\frac{m}{2\pi \nu}.
\ee
Thus, in the large-density limit $f_{core}$ converges to 1
\begin{equation}
    f_{core} \approx 1 - 2\sqrt{2}\sqrt{y} + O(y) \text{   for   }y\ll 1
\end{equation}
while in the small density limit 
\begin{equation}
   f_{core} \approx \frac{1}{4(y+1)^{2}}  \text{   for   }y\gg 1
\end{equation}
and the core constitutes a negligibly small fraction of the nodes.
The fraction of nodes belonging to the crossover region, where the in-degree increases with radial coordinate, is of the same order of magnitude, $f_{cross} \approx f_{core}(e^a - 1)$, where $a$ is, once again, the width of that region.

\begin{figure}
\includegraphics[width=15cm]{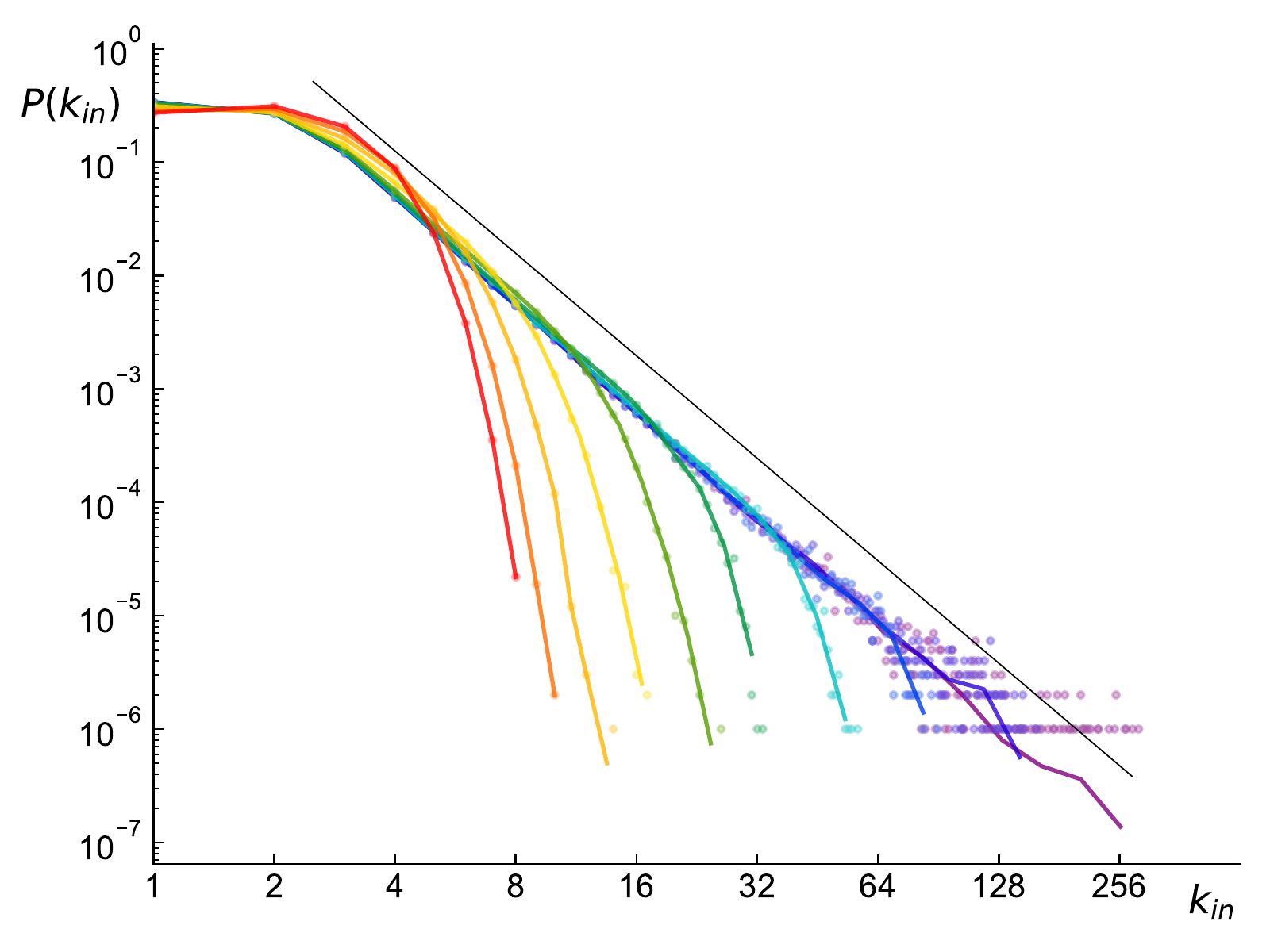}
\caption{In-degree distribution of the m-NN network with $m$=2 and $\nu$ changing from 0.5 (leftmost curve) to 0.001 (rightmost curve), $\nu$'s at adjacent curves differ by a factor of 2, curves are colored from red to purple in the rainbow order. The points are obtained by averaging over 100 independent networks of $N = 10000$ nodes for individual values of in-degree, continuous curves correspond to logarithmic binning with 25 bins per curve. The slope of the thin straight line is -3. }
\label{indegree_varnu}
\end{figure}

Thus, in terms of the overall distribution of in-degree there are two regimes. For $y=m/(2\pi \nu) \ll 1$ the network is dominated by the core where the in-degree distribution is approximately Poisson with average $m$. For $y \gg 1$ the network is dominated by the peripheral region where
\be
f(\zeta) =  e^{-\zeta},  \; \bar{k}_{in} (\zeta) \approx  \frac{m}{2} e^{\zeta/2},
\ee
where $0\leq \zeta \lesssim \ln 2(y+1)-a$. For each $\zeta$ true in-degree $k$ is narrowly distributed around $\bar{k}_{in} (\zeta)$. This allows to obtain the distribution of in-degree by replacing $k$ with $\bar{k}_{in}$ for each $\zeta$: 
\be
P_{in}(k) dk = f(\zeta) d\zeta \approx \left(\frac{2k}{m}\right)^{-2} d \left( 2\ln\frac{2k}{m} \right)= \frac{m^2}{2 k^3} dk; \;\; m<k<k_{in}^{\max},
\label{powerlaw}
\ee
where $k_{in}^{\max}$ is given by \eq{kinmax}. Thus, the in-degree distribution in the $y \gg 1$ is a truncated power law with exponent -3 spanning values from $k_{in}^{\min} \approx m$ to  $k_{in}^{\max} \approx 0.06 m^2/\nu$. \fig{indegree_varnu} shows the results of numerical simulations of the in-degree for networks with $m = 2$ and varying $\nu$ confirming this prediction.

\section{Structure beyond degree distribution: reciprocity}

In this section we go beyond the discussion of the degree distribution and discuss the simplest non-local structural characteristic of the network models introduced above. In directed networks the simplest nontrivial metric characterizing link-to-link interaction is called network reciprocity. It is related to the probability that, given that there is a link $i \to j$ there is also a link $j \to i$ and is conventionally defined\cite{Garlaschelli,Newman_recip} as
\begin{equation}
    r = \frac{N_{\rightleftarrows}}{N_b-N_{\rightleftarrows}}
\end{equation}
where $N_b$ is the total number of bonds in the network ($N_b = m N$ for the $m$-NN network) and $N_{\rightleftarrows}$ is the total number of pairs of bidirectional links. Note that in the usual definition of reciprocity \cite{Garlaschelli,Newman_recip} a bidirectional link is considered as a single link, while in our definition it corresponds to two links in different directions, thus the denominator of the reciprocity should be $N_b-N_{\rightleftarrows}$ in our notation. For the network with no bidirectional links $r=0$ , and $r=1$ if all links are bidirectional ($N_{\rightleftarrows}=N_b/2$ in this case). 

For undirected networks the simplest non-trivial metric of the relative position of the edges involves not 2 but 3 edges. It is the clustering coefficient of a node $c_i$ \cite{newman_book, barabasi_book}, defined as the fraction of node pairs adjacent to it which have a link between them, and the average clustering coefficient of a networks $C = N^{-1} \sum c_i$. To characterize clustering in directed networks the average clustering coefficient of the corresponding undirected network (i.e., the network obtained from the original one by replacing all directed links with undirected ones) is often used. Note, however, that such an approach leaves out the distribution of directed motifs in the original network, which is known to contain plenty of important information about its structure and function \cite{alon1,alon2}.

In geometrical random networks, where edge formation is based on distance between nodes (which is a symmetric function of node positions and obeys the triangular inequality), edges are spatially correlated. As a result both reciprocity and clustering coefficient remain of order 1 even in the limit of large networks. In what follows, we use the reciprocity as the main metric characterizing these correlations in the $m$-NN and VCR networks. We also provide numerical results for the average clustering coefficient of the corresponding undirected network. Study of the full distributions of directed motifs in these networks is of obvious interest but goes beyond the scope of the paper.

\subsection{Reciprocity of the VCR network}

We start with the simpler case of the network with varied connection radius. In this network a link $\ve x\to \ve y$ exists if distance $||\ve x - \ve y||$ is smaller than $r_m(\xi)$ while $\ve y\to \ve x$ exists if $||\ve x - \ve y||$ is smaller than $r_m(\zeta)$, where $\xi, \zeta$ are distances from points $\ve x, \ve y$ to the boundary of $O(R)$. Thus the link is bidirectional if 
\be
||\ve x - \ve y||\leq \min(r_m(\xi),r_m(\zeta)) = r_m(\max(\xi,\zeta))
\ee
, where we used that $r_m(\xi)$ is a monotonically decreasing function of $\xi$. Note that this means that outward-directed bonds (i.e., bonds for which $\zeta<\xi$) in VCR are always reciprocated, while inward-looking ones --- not necessarily so. Therefore, in this case reciprocity $r$ has also the meaning of the ratio of the numbers of outward and inward-directed bonds.

\begin{figure}
\includegraphics[width=17cm]{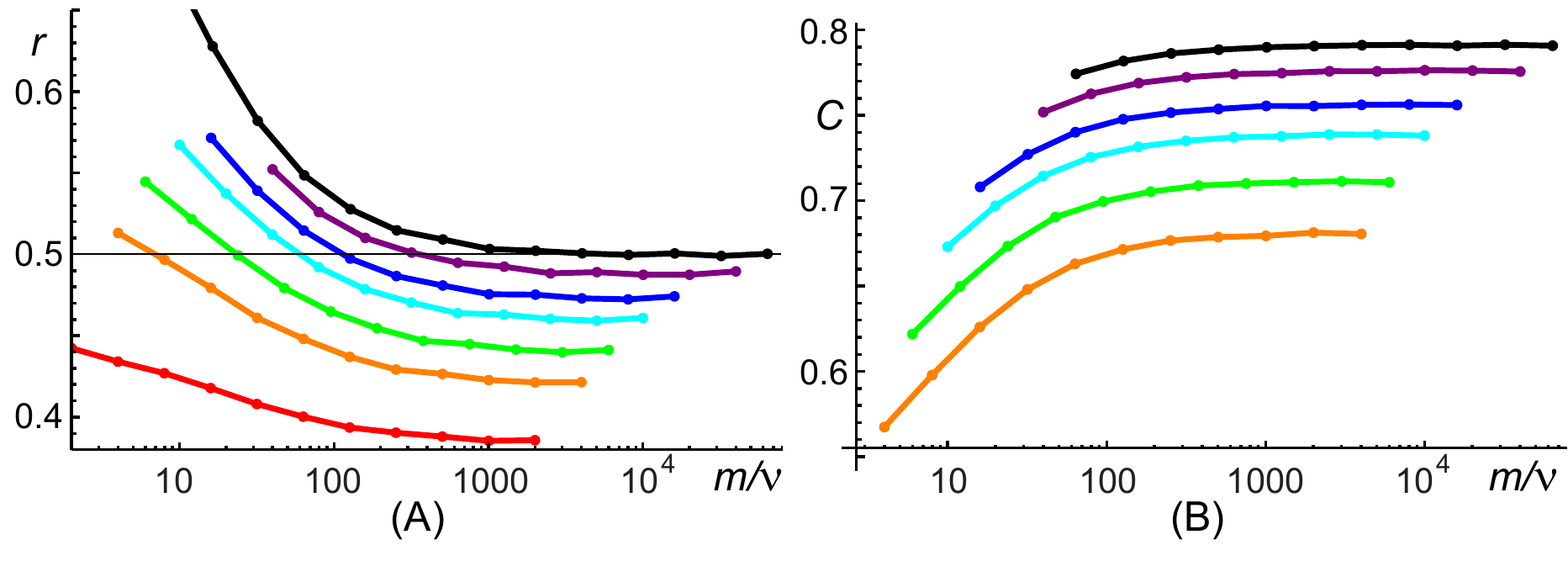}
\caption{(A) Network reciprocity $r$ and (B) average clustering coefficient $C$ of $m$-NN and VCR networks as functions of $m/\nu$. Numerical results for $m$-NN networks with (from bottom to top) $m = 1$ (red), 2 (orange), 3 (green), 5 (cyan), 8 (blue), 20 (purple) are shown, as well as for the VCR network (topmost, black, $m=32$). Dots are results of simulations average over 100 realizations of networks of $10^4$ nodes. The lines are but a guide to the eye. Note that VCR reciprocity converges to $r=1/2$ (thin horizontal line) for large $m/\nu$. Clustering coefficient for the $m$-NN network with $m=1$ is equal to 0.}
\label{reciprocity}
\end{figure}

The mean number of bidirectional links from node $\ve x$ can be written as
\begin{equation}
\begin{array}{rll}
    \bar{k}_{\rightleftarrows}(\ve x) &= & \bar{k}_{\rightleftarrows}(\xi) = \displaystyle \int_{O(R)} d\ve y \nu \Theta(r_m(\max(\xi,\zeta)) - ||\ve x - \ve y||) = \medskip \\   
    &=& \displaystyle \int_{\xi}^R \int_{-\pi}^\pi\nu \sinh (R-\zeta) d\zeta d\phi\; \Theta(r_m(\zeta) - ||\ve x - \ve y||) + \int_0^{\xi} \int_{-\pi}^\pi\nu \sinh (R-\zeta) d\zeta d\phi\; \Theta(r_m(\xi) - ||\ve x - \ve y||)
\end{array}
\end{equation}
where $\Theta(x)$ is the Heaviside theta-function. Using \eq{cosine} one gets in the large $R$ limit
\begin{equation}
\begin{array}{rll}
    \bar{k}_{\rightleftarrows}(\xi) &= & \displaystyle 2 \nu\int_{\xi}^{\zeta_{\max}} d\zeta  \sqrt{2[\cosh(r_m(\zeta))-1]e^{\xi-\zeta} - (e^{\xi-\zeta}-1)^2}  + \medskip \\
    & + & \displaystyle 2 \nu\int_{\zeta_{\min}}^{\xi} d\zeta  \sqrt{2[\cosh(r_m(\xi))-1]e^{\xi-\zeta} - (e^{\xi-\zeta}-1)^2}
\end{array}
\label{bidegree}
\end{equation}
where 
\begin{equation}
\left\{
    \begin{array}{rll}
      \zeta_{\min}& = \max[0, \xi-r_m(\xi)]  \medskip \\
      \zeta_{\max}& \text{ is the solution of  } \;\;\zeta_{\max} =\xi + r_m(\zeta_{\max}) 
    \end{array}
\right.
\label{k_bi}
\end{equation}
Given that $r_m(\xi)\geq r_m$ for all $\xi$ and $r_m(\xi) = r_m$ for $\xi\geq r_m$, this further simplifies to
\begin{equation}
\left\{
    \begin{array}{rll}
      \zeta_{\min}& = \max[0, \xi-r_m]  \medskip \\
      \zeta_{\max}& = \xi + r_m 
    \end{array}
\right.
\end{equation}
In \fig{mf_indegree} the results of numerical integration of \eq{bidegree} for $m/\nu=1, 100, 10^4$ are shown with blue lines. For $\xi>r_m$ $\bar{k}_{\rightleftarrows}(\xi)\equiv m$, i.e. all out-bonds are simultaneously in-bonds. For smaller $\xi$ the number of bi-directional bonds is smaller than $m$, obviously
\be
\min(\bar{k}_{in} (\xi),m)\geq \bar{k}_{\rightleftarrows}(\xi).
\ee
Reciprocity can be easily expressed in terms of mean number of bidirectional bonds $\bar{k}_{\rightleftarrows}$. Indeed, the expected total number of such bonds is
\be
N_{\rightleftarrows}=\frac{N}{2}\int_0^{\infty} f(\xi) \bar{k}_{\rightleftarrows}(\xi) d\xi
\ee 
where $N$ is the total number of nodes in the network, $f(\xi)d \xi$ is the fraction of nodes with spatial coordinate $\xi$, $f(\xi) = \exp(-\xi)$ for large $R$, see \eq{fzeta}. In the small density limit $m/\nu \gg 1$ one gets in the leading order in $\mu$
\be
\cosh r_m(\xi) \approx \frac{1}{32} \left(\frac{m}{\nu}\right)^2 e^{-\xi}
\ee
Substituting this into \eq{k_bi} one gets in the leading order in $m/\nu$
\begin{equation}
    N_{\rightleftarrows}=\frac{N m}{4} \left( \int_0^{\infty}e^{-\xi} d\xi \int_0^{\xi}e^{-\zeta/2}d\zeta +\int_0^{\infty}e^{-\xi/2}d\xi\int_{\xi}^{\infty}e^{-\zeta} d\zeta \right) = \frac{N m}{3}
\end{equation}
Thus, $r$ converges to $1/2$ for large densities. In turn, in the small $m/\nu$ regime the network is dominated by the core where all bonds are bidirectional, and $r$ converges to 1 for $m/\nu \to 0$. This predictions are confirmed by numerical simulations of large VCR networks as shown in \fig{reciprocity}.

\subsection{Reciprocity of the $m$-NN network}

Consider now the reciprocity of the $m$-NN networks. Most notably, it is clear that reciprocity of the $m$-NN network is smaller than that of the corresponding VCR network. In particular, in the VCR all bonds far from the boundary are bidirectional and all outward-directed bonds are reciprocated, while in the $m$-NN they are not. Indeed, for example, for VCR the condition to connect two nodes in the core is $||\ve x - \ve y|| < r_m$ which has a $\ve x \rightleftarrows \ve y$ symmetry, while the relation ``$x$ is $m$-th nearest neighbor of $y$'' is not. However, one expects the reciprocity of the $m$-NN networks to approach that of the VCR network in the limit of $m \to \infty, m/\nu = const$. 

\begin{figure}
\includegraphics[width=5cm]{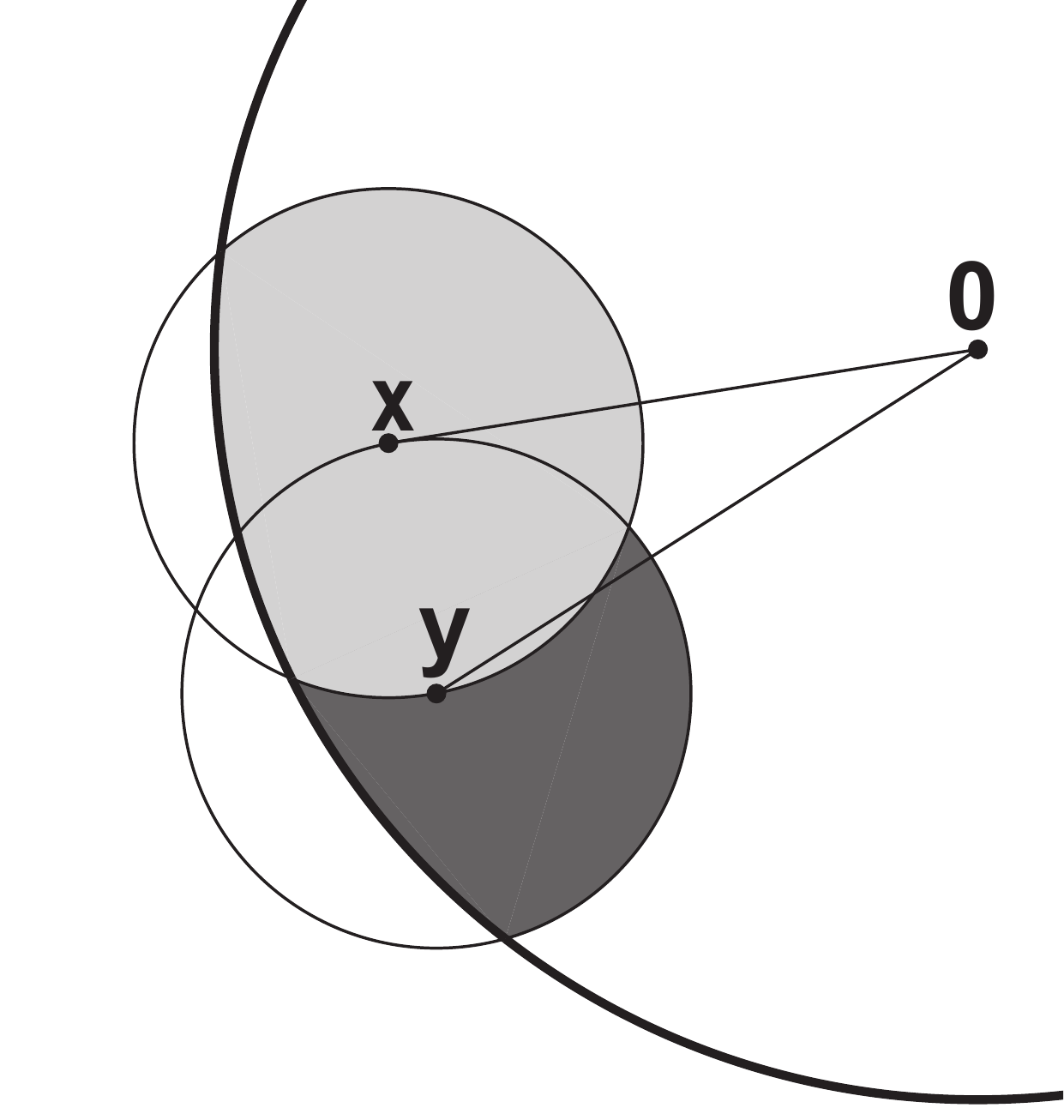}
\caption{A sketch illustrating the calculation of the probability of a bidirectional link in a $1$-NN network. Thick line is the boundary of $O(R)$, the thin circles have radius $||\ve x - \ve y||$. $\ve y$ is the nearest neighbour of $\ve x$ if there are no nodes in the light-grey area, $\ve y$ is simultaneously the nearest neighbour of $\ve x$ if additionally there are no nodes in the dark-grey area.}
\label{recip_sketch}
\end{figure}

We have not been able to calculate reciprocity explicitly in the general case of an $m$-NN network. However, here we provide some estimates for the case of 1-NN network. Given that VCR network corresponds to the $m \to \infty$ limit, it is natural to expect that reciprocity for arbitrary $m$ to lie between these two limiting cases (these expectations are confirmed by numerical simulations, see \fig{reciprocity}).

Let $\ve x$ be a position of a node and $\ve y$ -- the position of its nearest neighbor at distance $r = ||\ve x- \ve y||$. This means that there are no other nodes inside the light grey area in the sketch \fig{recip_sketch}. In turn, for node at $\ve x$ to be the nearest neighbour of the node at $\ve y$ there should be no nodes closer to $\ve y$ than $r$, i.e., there additionally should be no nodes in the dark gray figure in the sketch. The area of this figure can be written formally as
\begin{equation}
    A_{\rightleftarrows} (\ve x, \ve y, R) =\int \Theta (R-z) \Theta(r-||\ve y - \ve z||) \Theta (||\ve x - \ve z|| -r) d \ve z,
\label{A_2}
\end{equation}
where we keep using the notation $||\ve z|| = z$. The probability of there being no nodes in this area is $\exp(-\nu A_{\rightleftarrows})$ and the overall probability that bond from $\ve x$ to its nearest neighbor is bidirectional is obtained by integrating over possible position of the node $\ve y$ 
\begin{equation}
    p_{\rightleftarrows}(\ve x,\nu, R) =\int_{O(R)} \nu d\ve y \; \pi_1(\ve x, \ve y, \nu, R) \exp(-\nu A_{\rightleftarrows} (\ve x, \ve y, R)) =\int_{O(R)} \nu d\ve y \exp\left[-\nu(A(\ve x, \ve y, R)+ A_{\rightleftarrows} (\ve x, \ve y, R)) \right]
    \label{p_2}
\end{equation}
where $\pi_1$ is given by \eq{pi_circle}. Total number of bidirectional bonds $N_{\rightleftarrows}$ can be integrating \eq{p_2} over the spatial distribution of the nodes. Notably, 
\begin{equation}
    \int \pi_1 d \ve y = \int_{O(R)} \nu d\ve y \exp(-\nu(A(\ve x, \ve y, R)) = 1-O(e^{-R})
\end{equation}
(i.e., every node has a single nearest neighbour provided the network has at least two nodes), so $p_{\rightleftarrows}<1$ as soon as $A_{\rightleftarrows}>0$. Explicit calculation of \eq{A_2} is very cumbersome. Here we limit ourselves to the study of the core of the network where $\max(x,y)\leq R-||\ve x - \ve y||$ and one can therefore omit the first $\Theta$-function in \eq{A_2}. In this case explicit calculation gives
\be
A_{\rightleftarrows} (||\ve x - \ve y||=r) = 4\cosh r \arctan\left(\frac {\cosh r}{\sqrt{1+2\cosh r}} \right) + 4 \arctan \sqrt{1+2\cosh r} - 2\pi
\ee
which asymptotically behaves as
\begin{equation}
    \left\{
    \begin{array}{ll}
     \displaystyle \frac{2\pi + 3\sqrt{3}}{6}r^2 + O(r^4)   & \text{  for  } r\ll 1, \medskip \\
     \displaystyle 2\pi \cosh r - 8 \cosh (r/2) + O(e^{-r})    & \text{  for  } r\gg 1,
    \end{array}
    \right.
\end{equation}
and thus the fraction of bidirectional bonds in the core of the network is 
\be
\begin{array}{rll}
P_{\rightleftarrows}^{(core)}&= & \displaystyle \pi\int_0^{\infty} \nu \sinh r dr \exp (-\nu (A(r)+A_{\rightleftarrows}(r))) =\frac{1}{2}\int_0^{1}  e^ {-\nu A_{\rightleftarrows}} d \left(e^ {-\nu A}\right) \medskip \\
r^{(core)} & = & \displaystyle \left( \frac{1}{P_{\rightleftarrows}^{(core)}} -1 \right)^{-1} 
\end{array}
\label{rcore}
\ee
This integral is easy to calculate numerically, one gets the following limits for reciprocity in the core for large and small $\nu$
\be
r^{(core)}= \left\{ 
\begin{array}{ll}
    \displaystyle \frac{3\pi}{5\pi + 3\sqrt{3}} \approx 0.4509... & \text{ for } \nu \to \infty, \medskip\\
    \displaystyle \frac{1}{3} & \text{ for } \nu \to 0.
\end{array}
\right.
\ee
For large $\nu$ the network is dominated by the core and one expects
$r\approx r^{(core)}$ which is confirmed by numerical results shown in \fig{reciprocity}A (the red line corresponds to $m=1$). In the boundary-dominated regime the reciprocity is larger than the low-density core value of $1/3$ but still smaller than the reciprocity of the small $\nu$ network.

\subsection{Regulating reciprocity with changing temperature}

Reciprocity varies widely in the experimentally observed directed networks\cite{Garlaschelli,boguna_dir}. For example, citation networks\cite{citations} and mathematical proof networks \cite{dedeo} have zero or essentially zero reciprocity. Trade networks have reciprocity close to 1 \cite{Garlaschelli}. However, large classes of networks have intermediate reciprocities. For example, free association networks we mentioned in the introduction as a proxy of a more general reference network typically have reciprocity of order 0.1 \cite{Garlaschelli, valba} which is, on the one hand side, much larger than one would expect in a random network, but, on the other hand, significantly smaller than what is expected in the $m$-NN network with similar out-degree ($m \approx 30$).

This gives rise to a question whether it is possible to modify the definition of $m$-NN and VCR models in a way that the value of reciprocity becomes adjustable. This can be done by introducing some form of stochastisity in the formation of the links in the network. Start once again with the VCR. Recall that there is a bond from a node at $\ve x$ to a node in $\ve y$ if and only if $||\ve x - \ve y||< r_m(\ve x)$, where $r_m(\ve x)$ is the solution of \eq{r_mx}, i.e. putting it more formally, the probability that a bond is formed between two points is 
\begin{equation}
    p_{\ve x \to \ve y} =  P(||\ve x - \ve y||, r_m(\ve x)) = \Theta(r_m (\ve x) - ||\ve x - \ve y||)
\end{equation}
where $r_m(\ve x)$ is determined by 
\begin{equation}
    \int_{O(r)}P(||\ve x - \ve y||, r_m(\ve x)) d\ve y = m/\nu
\label{normalize}
\end{equation}
Formulated like this, it is clear that VCR model can be naturally generalized to an arbitrary non-negative kernel $P(||\ve x - \ve y||, \sigma(\ve x))$ describing the probability of bond formation, $\sigma$ being some set of governing parameters. If these parameters are chosen according to normalization rule \eq{normalize}, the resulting network has a position-independent average out-degree $m$. One possible natural choice is a Gaussian kernel
\begin{equation}
    P(||\ve x - \ve y||, \sigma(\ve x))=\exp\left( - \frac{||\ve x - \ve y||^2}{\sigma(\ve x)^2}\right),
\end{equation}
which is used, for example as a measure of point similarity in t-SNE \cite{tsne} and to estimate the probability of bond formation in polymer physics (see, e.g., \cite{lgk,erukh1}). It is, however, not  convenient for our purposes since it has only a single governing parameter $\sigma(\ve x)$, which is to be fixed by \eq{normalize}. As a result, it is not possible to regulate out-degree and reciprocity independently in this case. We consider instead the Fermi-Dirac kernel
\begin{equation}
     P(||\ve x - \ve y||, r_m, T) =  \left[1+ \exp\left(\frac{||\ve x - \ve y||-r_m(\ve x, T)}{T}\right)\right]^{-1},
    \label{fermi_dirac}
\end{equation}
which was suggested in a similar context in \cite{krioukov2} (note also that it is a natural regularization of the truncated exponential kernel used for the probability of nearest-neighbour bonds in \cite{umap}). The kernel \eq{fermi_dirac} has two parameters - temperature of bond formation $T>0$ and typical connection radius $r_m(\ve x, T)$, and it converges to the VCR in the limit $T\to \infty$ (the normalizaion condition \eq{normalize} converges to \eq{r_mx} in this case). In the model with the Fermi-Dirac kernel the formation of bonds is stochastic, and one expects the reciprocity of the network to decrease with increasing $T$ without changing of other essential properties of the networks such as the in- and out-degree distributions. 

In order to understand how the structure of the network depends on $T$, rewrite the equation \eq{normalize} in the form
\begin{equation}
    \int dr \frac{\partial A(r, x, R)}{\partial r} \left[1+ \exp\left(\frac{r-r_m(\ve x, T)}{T}\right)\right]^{-1}  = m/\nu
\label{normalize_FD}
\end{equation}
where $A(r,x,R)$ is collectively defined by \eq{asmall}-\eq{over_integral}, and study the large $r$ tail of the integrand in the left hand side. Importantly, for large radii ($e^r \gg 1$) the area $A(r,x,R)$ grows exponentially, so that 
\begin{equation}
\frac{dA(r,x,R)}{dr} \sim \left\{ 
\begin{array}{ll}
\exp r\ \ \ & \text{for } r<R-x \medskip \\
\exp[(r+R-x)/2] \ \ & \text{for } R-x<r<R+x
\end{array}
\right.
\end{equation}
(see \eq{Approx_A},\eq{Approx_A2}). In particular, for $x$ in the center and on the boundary of the disk we get, respectively,
\begin{equation}
    dA(r,0,R)/dr \sim \exp r, \ \ dA(r,R,R)/dr \sim \exp (r/2)
\end{equation}
The large $r$ behavior of the integrand in \eq{normalize_FD} is controlled by the competition between the exponential growth of $dA(r,x,R)/dr$ and the exponentially decaying factor $\exp[-r/T]$, coming from the Fermi-Dirac kernel. If the product of these two factors is decaying for large $r$, the typical bond length is close to $r_m$ as it is in the zero-temperature case (we call this case ``the short bonds regime''). In turn, if the product is growing for large $r$ the typical bond length decouples from $r_m$ and is, in fact close to the maximal possible value $R+x$ (we call this ``the long bonds regime", $r_m$ in this case plays a role of $R$-dependent normalization constant). 

Thus, depending on temperature, one expects the following three regimes.

If $T<1$ all bonds are short, with their length controlled by $r_m(\ve x, T)$. For example, for bonds starting in the center of the disk it is easy to estimate the mean bond length by approximating the Fermi-Dirac factor \eq{fermi_dirac} with 
\begin{equation}
     P(r, r_m, T) = \left\{
     \begin{array}{ll}
     1 \ \ \ \ &\text{for } r<r_m \medskip \\
     \exp(-(r-r_m)/T)\ \ &\text{for } r>r_m
     \end{array}
     \right.
    \label{fermi_dirac_approx}
\end{equation}
which gives
\be
\langle l \rangle \approx r_m - 1 + T/(1-T).
\ee
As a result, the structure of the network in this regime is  qualitatively similar to that at $T=0$: the network has a core region, where properties (in-degree, reciprocity, etc) do not ``feel'' the boundary of the disk, and peripheral region with roughly exponential dependence of the in-degree on the distance to the boundary.

For $T>2$ all bonds are long, the nodes predominantly connect to nodes which are very far from them: there are exponentially many such nodes, and Fermi-Dirac decay is not strong enough to prevent formation of the long bonds. As a result, in the thermodynamic limit bonds become completely uncorrelated and the reciprocity converges to zero.

The most interesting situation arises for $T\in (1,2)$. In this case nodes close to the center of the disk predominantly are sources of long bonds (i.e., they typically connect not to their neighbours but to random nodes close to the periphery of the disk), while peripheral nodes are sources of short bonds, connecting them mostly to their neighbours. As a result, network does not have a core-periphery structure anymore. However, since the global properties of the network are mostly controlled by peripheral nodes, the overall in-degree remains a power law.

\begin{figure}
\includegraphics[width=17cm]{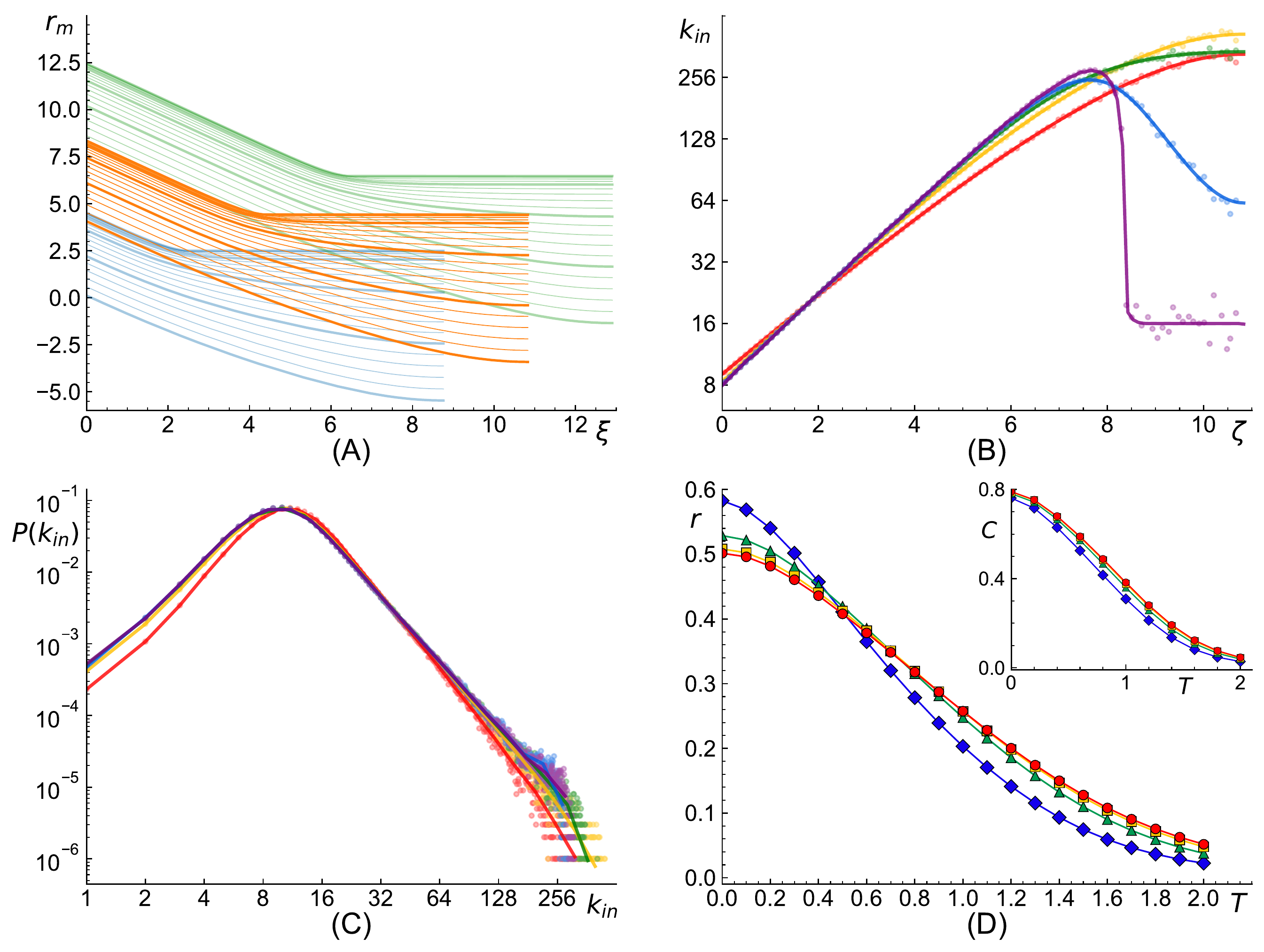}
\caption{Structural properties of temperature-dependent VCR networks. (A) the dependence of $r_m(\xi,T)$ defined by \eq{normalize_FD} on the distance from the boundary $\xi$ for networks of $N=10000$ nodes with $m$=16 and (from bottom to top) $\nu = 1/2$ (blue), $1/16$ (orange), $1/128$ (green) (the corresponding disk radii are $R\approx 8.76, 10.83, 12.90$, respectively) and various temperatures. In each series of curves temperature grows from $T=0$ (top curve) to $T=2$ (bottom curve) with step 0.1. (B) the dependence of average in-degree of the node on its distance $\xi$ from the boundary; dots are results for simulated networks of $N=10000$ nodes with $m$=16, $\nu =1/2$, $R\approx 10.83$ and temperatures $T=0., 0.5, 1, 1.5$ and 2, colored from purple to red in the rainbow order; curves are results of numerical integration of \eq{TVCR_in}, points correspond to average of 100 independent model networks; (C) global in-degree distribution for the same networks as in panel (B); (D) reciprocity and (in the inset) average clustering coefficient as functions of temperature for networks with $N=10000, m =16$ and $\nu=1/2$ (blue diamonds), $1/8$ (green triangles), $1/32$ (orange squares) and $1/128$ (red circles). }
\label{TVCR}
\end{figure}

In \fig{TVCR} we present the results of numerical simulations of temperature-dependent VCR with the Fermi-Dirac kernel. \fig{TVCR} (A) shows the behavior of the connection radius $r_m(\xi,T)$ for different temperatures and $m/\nu$ and fixed $N=10000$. It is seen that for low temperatures (upper curves in the series) the behavior is qualitatively similar to that of zero-temperature VCR: there is clear distinction between the core of the network where $r_m$ is independent of the coordinate and periphery with linear dependence of $r_m$ on $\xi$, However, the core shrinks with growing temperature and ceases to exist for $T>1$. In the $T>1$ regime $r_m$ actually can become negative. Note, however, that in this regime the length of bonds is close to $R+x=2R-\xi$ and $r_m$ plays just a role of normalization constant in \eq{normalize_FD}.

\fig{TVCR} (B) shows the behavior of the average in-degree of a node as a function of coordinate,
\be
k_{in}(\zeta) = \int_{O(R)}  P\left(||\ve x - \ve y||, r_m (\ve x, T), T\right) \nu d\ve x
\label{TVCR_in}
\ee
where $P(r,r_m,T)$ is given by \eq{fermi_dirac} and the integral, although formally a function of the target node position $\ve y$ in fact depends only on the distance to the boundary $\zeta = R-y$. Interestingly, the plateau of the in-degree distribution in the core of the network exists only up to $T=0.5$. However, up to $T=1$ the in-degree dependence on $\zeta$ is non-monotonic, exponential increase for small $\zeta$ is followed by decrease at larger $\zeta$ similarly to the low-temperature limit. For $T>1$ the in-degree keeps slowly increasing up to the center of the disk. However, given that the majority of the nodes are located in the peripheral region where exponentially increasing $k_{in}(\zeta)$ dependence persists for all temperatures $T\in [0,2)$, it is not surprising that the global in-degree distribution remains a power law with exponent -3 (see \fig{TVCR} (C)). In fact, erosion of the core regime with growing temperature leads to the cut-off of the power law being shifted to higher in-degrees.

Finally, \fig{TVCR} (D) shows the numerical results for dependence of reciprocity and clustering coefficient on temperature for various $m/\nu$ and $N=10000$. As expected, both decrease with increasing temperature. The fact that reciprocity at $T=2$ remains small but finite is a finite-size effect: it is easy to show that $r(T=2)$ should decrease logarithmically with the size of the network.

To conclude, note that it is possible to generalize the idea of stochastic temperature-controlled bond formation to the case of $m$-NN networks in this case the probability of forming a bond from node $i$ to node $j$ can be defined as
\begin{equation}
    p_{i \to j} = \left[1+\exp\left(\frac{m_{ij}-\mu(T)}{T} \right)\right]^{-1}
\end{equation}
where $m_{ij}$ is the order of $j$ in the list of nearest neighbors of $i$, as defined above \eq{kin_mnn} ($m_{ij}$ =1 for the nearest neighbour of $i$, 2 for the next-nearest neighbor, etc.), and $\mu(T)$ is defined in a way to make the mean out-degree fixed:
\begin{equation}
    \sum_{k=1}^{\infty} \left[1+\exp\left(\frac{k-\mu(\ve x,T)}{T} \right)\right]^{-1} = m
\end{equation}
or in some other similar way. 

We expect the qualitative behavior of such a network to be similar to that of the T-VCR: the reciprocity of the network to decrease with growing temperature while in-degree distribution remaining a truncated power-law.

\section{Variation of the exponent of the in-degree distribution}

The exponent of the in-degrees distribution in the power-law (boundary-dominated) regime is fixed at -3 for both VCR and $m$-NN models. Here we briefly discuss how to relax this constraint. To do this, following \cite{krioukov2}, we introduce a ``quasi-uniform'' distribution of nodes, so that the density of nodes at distance $r$ from the origin behaves as
\begin{equation}
    \nu(r) = \nu\; \frac{\sinh \alpha r }{\sinh r}
\label{quazi}
\end{equation}
with some constant parameter $\alpha$. This distribution of nodes breaks the translation invariance of the properties in the core of the network (e.g., the mean in-degree becomes slightly coordinate-dependent) but in the periphery regime the resulting behavior is quite simple: the fraction of nodes with a given distance to the boundary becomes
\begin{equation}
    f_{\alpha}(\zeta) d\zeta = e^{-\alpha \zeta} d\zeta
\end{equation}
instead of \eq{fzeta}, which, after substitution into \eq{powerlaw} gives 
\begin{equation}
    P_{\alpha}(k_{in}) \sim k_{in}^{-1-2\alpha}, \ \  F_{\alpha}(k_{in}) \sim k_{in}^{-2\alpha},
    \label{quasiu_exp}
\end{equation}
for the pdf an the cumulative degree distribution, respectively (compare \cite{krioukov2}) and one needs $\alpha \leq 2$ for the whole network to stay in the boundary-dominated regime in the thermodynamic limit.

\begin{figure}
\includegraphics[width=17cm]{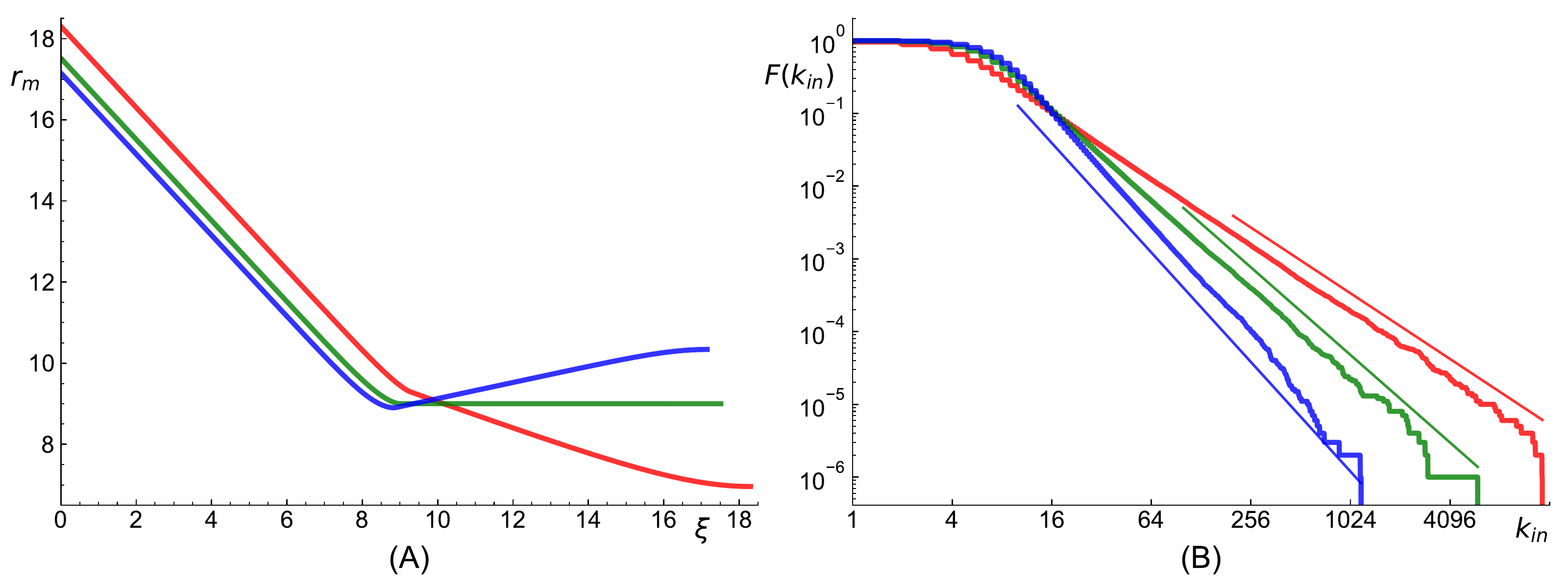}
\caption{(A) Connection radius $r_m$ as a function of the distance $\xi$ of the source node from the boundary for networks with quasiuniform distribution of nodes \eqref{quazi}, $\alpha = 0.75$ (red, the curve decreases inside the core, i.e. for large $\xi$), 1 (green, uniform case,  constant in the core), 1.25 (blue, increases in the core); numerical results for $m=10, N=50000$ and radius of disk $R$ defined by equation $r_m(\xi=0)=R$. It is seen that quasiuniform distribution violates the translation invariance in the core but in the periphery the dependence $r_m(\xi) = r_m(0) -\xi$ still holds; (B) cumulative in-degree distributions of the same networks with (from top to bottom) $\alpha = 0.75$ (red), 1 (green, uniform case), 1.25 (blue), the corresponding straight lines indicate the corresponding power law behavior given by \eqref{quasiu_exp}. }
\label{fig_alpha}
\end{figure}

In \fig{fig_alpha} we show the numerical results for the connection radius $r_m(\xi)$ and for the cumulative in-degree distribution of the VCR networks with various $\alpha$, showing that for networks with small enough node density, despite the violation of the translation invariance in the core, the in-degree distribution indeed does have a power law tail \eqref{quasiu_exp}.

\section{Generating networks with given properties using the nearest-neighbour models}

In this section we briefly discuss how to use network models introduced above to generate directed networks with given properties. Recall that out-degree in the models considered here is fixed by construction,
\be
P_{out}(k) = \left\{
\begin{array}{ll}
   \delta_{k,m}  & \text{for the $m$-NN model}; \\
    m^k e^{-m}/k! & \text{for the VCR model},
\end{array}
\right.
\label{out}
\ee
while other properties of the networks, such as the slope and width of the in-degree distribution and the reciprocity can be tuned by adjusting the parameters of the models.

As an example, consider constructing a network whose in-degree has a purely power-law tail with slope -3. We also demand that the network has a given average degree $m$, number of nodes $N$ and reciprocity $r$. For definiteness, let us construct it based on the VCR model.  

In the large $N$ limit the in-degree distribution of a VCR network is, for any given $\nu$, a truncated power law. However, for any final $N$ it is possible to choose density of nodes in a way that only the power-law part of the distribution is observed. Indeed, the truncation of the power-law is caused by the existence of the core part of the network where the expected degree does not depend exponentially on the distance to the boundary. Thus, if one chooses the density of points in a way that the width of the boundary region becomes equal to the radius of the disk, the truncated part of the distribution ceases to exist. Using \eqref{circle} for the area of hyperbolic circle and \eqref{rmx_approximate} for the connection radius at the boundary $r_m(0)$ one gets
\be
\left\{
\begin{array}{rll}
N &= &2\pi\nu (\cosh R - 1) \approx \pi \nu e^R \medskip \\
r_m(0) & = & \displaystyle 2 \ln\left[\frac{m}{4\nu} +\frac{\pi}{2}\right] \medskip \\
r_m(0) & = & R
\end{array} 
\right.
\ \ \rightarrow \ \ \ \ \nu_0\approx \frac{\pi}{16} \frac{m^2}{N},\ \  R_0\approx 2\ln\left(\frac{4N}{\pi m}\right)
\label{nu_est}
\ee
where $N/m \gg 1$ is assumed. Equation \eqref{nu_est} gives the value of $R_0$ ($\nu_0$) for which the in-degree distribution of the resulting networks is  the closest to a true power law. If $R<R_0$ is chosen (which means that $\nu$ is larger than $\nu_0$), the network has a core which is unreachable from the boundary and has bulk properties, including position-independent average in-degree $m$, while if $R>R_0$ is chosen an "inverse core" of radius $r_m(0)-R$ is formed in the center of the network, the nodes in this inverse core get connected to all peripheral nodes, attaining in-degree of order $N$. Note, however, that if radius is close to $R_0$ the actual number of nodes in core (reverse core) is small (it is of order $N \exp(-r_m(0))$ for the core and $N \exp(-2R+r_m(0))$ for the reverse core).


The second step of the network construction is to distribute $N$ points at random in the hyperbolic disk of the chosen radius $R_0$ and calculate the distances between them. It can be done in a standard way: the coordinates $(\rho_i,\theta_i)$ of the nodes ($i=1,\dots,N$) are chosen independently at random from probability distributions
\begin{equation}
\left\{
\begin{array}{ll}
    p_{rad}(\rho) = \sinh \rho/(\cosh R_0 - 1) & \rho\in[0,R_0] \\
    p_{ang}(\theta) = 1/2\pi & \theta \in[0,2\pi)
\end{array}    
\right.
\label{densities}
\end{equation}
The distance $d_{ij}$ between each pair of points can be calculated according to the hyperbolic cosine theorem \eq{cosine} as
\begin{equation}
    d_{ij} = \text{arccosh}\left(\cosh \rho_i \cosh \rho_j - \sinh \rho_i \sinh \rho_j \cos (\theta_i -\theta_j)\right).
\end{equation}

Third step is to determine the value of temperature $T$ corresponding to the desired value of reciprocity $r$. Note (see \fig{TVCR}D) that for large values of $m/\nu$ the $r/T$ dependence rapidly converges to the master curve, so for $m/\nu \gg 10^2$ for practical purposes the curve is universal (compare red and orange curves in \fig{TVCR}D). The limiting curve for $N=...$ is tabulated in the Appendix. There is a slow (logarithmic) dependence on $N$ of this limiting curve. However, a first estimate of $T$ based on the tabulated values turns out to be rather precise (see results of computer simulation below), and it is easy to improve the estimate iteratively, if needed. 

Fourth, one should find the dependence $r_m(\xi)$ for the chosen values of $m,\nu,R,T$. In the first approximation one can (compare \fig{TVCR}A) use a simple piece-wise linear fit
\be
r_m(\xi) = \max (r_m(0)-\xi, r_m(R))
\label{linear}
\ee
where the values at the boundary $r_m(0)$ and at the center of the disk $r_m(R)$ can be approximated by solutions of the following equations
\be
\begin{array}{rll}
_2F_1 \left( 1, T/2; 1+T/2; -\exp[(2R-r_m(0))/T]\right) &= &\displaystyle \frac{\pi m}{4N}, \medskip \\ 
_2F_1 \left( 1, T; 1+T; -\exp[(R-r_m(R))/T]\right) &= &\displaystyle \frac{m}{N},
\end{array}
\label{hypergeom}
\ee
where $_2F_1(a,b;c;d)$ is the ordinary hypergeometric function, and \eqref{hypergeom} is obtained from \eqref{normalize_FD} by replacing hyperbolic functions with respective exponents in the expressions for $A(r,x,R)$. Finally, a directed bond between each pair of points is formed independently at random with probability given by \eqref{fermi_dirac}. 

\begin{figure}
\includegraphics[width=17cm]{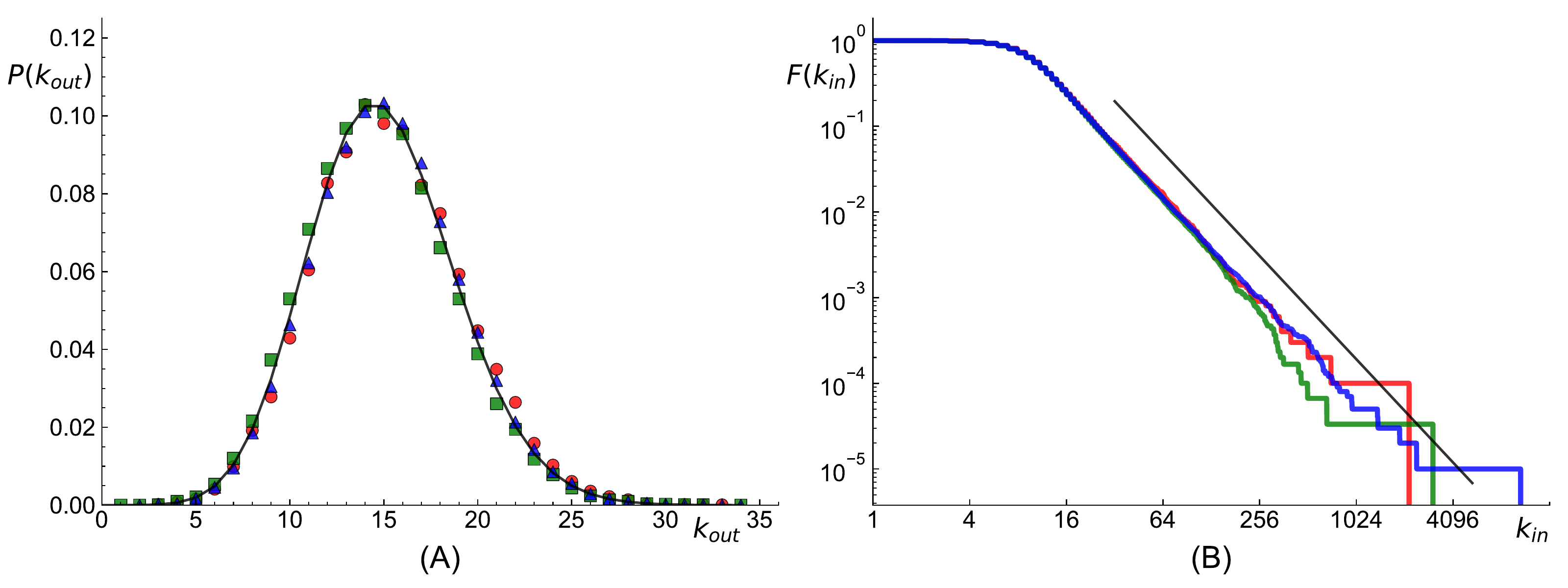}
\caption{(A) Out-degree distribution of models with $m=15, r=0.2$ and $N=10^4$ (red circles), $3\times 10^4$ (green squares) and $10^5$ (blue triangles) compared with the corresponding Poisson distribution (solid line); (B) cumulative in-degree distribution for the same networks, the straight line has a slope -2 and is a guide for the eye.}
\label{koutmod}
\end{figure}

We have evaluated this procedure for 27 different combinations of parameters $N = 10^4, 3\times10^4, 10^5$, $m=5, 15, 40$ and $r=0.1,0.2,0.4$. The comparison between the seed values of the parameters and the corresponding parameters in the constructed networks are provided in the Appendix. The results are mostly satisfactory: parameters of the constructed models are within 1-2\% of the desired seed values. Two minor problems are (i) notable discrepancy between constructed and seed value of resiprocity for small value of seed $r$, and (ii) a small systematic error in the average degree. Both effects are especially notable for smaller $N$ and seem to decline with growing $N$. The discrepancy in $r$ is due to the aforementioned non-universal behavior of the $r(T)$ curve for large $T$ and can be corrected by iteratively changing the value of temperature used in simulations. The discrepancy in the average degree is due to the fact replacement of the true behavior of $r_m(\xi)$ with a piecewise linear approximation \eqref{linear} and, if needed, can be alleviated by using a more sophisticated approximation for $r_m(\xi)$, e.g., 
\begin{equation}
    r_m (\xi) = r_m(R) + (|X| - X + a \exp(-|X|/a))/2,\ \ X = \xi - r_m(0) - r_m(R)
    \label{expanzats}
\end{equation}
with a properly adjusted numerical parameter $a$ (note that for $a\to 0$ \eqref{expanzats} converges to \eqref{linear}). Moreover, in \fig{koutmod} we show the resulting in- and out-degree distributions for several of the networks proving that out- and in-degree distributions of the generated networks are indeed Poison and power law with exponent 3, respectively.

\section{Discussion}

In this paper we studied the properties of the nearest neighbor model and a dual model with varied connection radius on a hyperbolic disk. We have shown that the properties of these networks change dramatically with the change of dimensionless (per unit of inverse curvature squared) density of nodes. 

If density of nodes is large (effective space curvature is small) the  resulting networks are similar to the nearest-neighbour networks in Euclidean space: both out- and in-degree are narrowly distributed around their mean value $m$ and the reciprocity of the network is very large, between $r=(5/3 + \sqrt{3}/\pi)^{-1} \approx 0.4509$ for $1$-NN networks and $r=1$ for the VCR networks, which correspond to the $m \to \infty$ limit of $m$-NN. 

In turn, if density of nodes is small (conversely, space curvature is large) the networks can be thought of as directed analogues of undirected hyperbolic networks studied in \cite{krioukov2,krioukov3}. Let us point out the important generality of this result: although $m$-NN and VCR networks are 3-parametric ($m,\nu, R$) classes of networks, in the limit of large network size their behavior is remarkably universal. Indeed, the in-degree distribution is always a power law with exponent $-3$, for $k_{in}/m$ from 1 to $m/\nu$, i.e. the only relevant parameter is $m/\nu$, which controls the position of the cut-off. Reciprocities of large networks converge to finite values for $\nu \to 0$. For VCR this value is $m$-independent, for $m$-NN it weakly depends on $m$ and converges to the VCR value for $m\to \infty$. In Table I we summarize the properties of all the networks studied in this paper in the limit $N\to\infty$ and for large but finite $m/\nu$.

This behavior comes from the peculiar core-periphery structure of the network, where in the core network properties are translation invariant and qualitatively similar to those of Euclidean nearest-neighbour networks, while in the boundary region the degree and local reciprocity are strongly position dependent, which, after integration over the spatial distribution of nodes gives rise to the scale-free behavior. In the thermodynamic limit $R\to\infty, \nu = const$ both core and periphery regions are present for all $m$ and $\nu$. However, distribution of the nodes between two regions depends on $m/\nu$ giving rise to core-dominated behavior for large densities and boundary-dominated behavior for small densities.

\begin{table}[ht!]
\centering
\newcommand{\cell}[2][2cm]{\parbox[m]{#1}{\vspace{5pt}#2\vspace{5pt}}}
\begin{tabular}{ | m{2cm} || m{2cm} | m{2cm} | m{2cm} | m{2cm} |}
\hline
    \cell{Model} & \cell{m-NN}  & \cell{VCR} & \cell{temperature-dependent VCR} & \cell{m-NN/VCR with quazi-uniform distribution} \\ \hline \hline

    \cell{parameters} & \cell{$m,\,\nu,\,R$} & \cell{$m,\,\nu,\,R$} & \cell{\vspace{5pt}$m,\,\nu,\,R,\,T$} & \cell{$m,\,\nu,\,R,\,\alpha$} \\ \hline

    \cell{out-degree distribution} & \cell{delta-functional} & \cell{Poisson} & \cell{Poisson} & \cell{delta/Poisson} \\ \hline

    \cell{in-degree exponent} & \cell{$-3$} & \cell{$-3$} & \cell{$-3$} & \cell{-1-2$\alpha$} \\ \hline

    \cell{in-degree cut-off $k_{max}/m$} & \cell{$\sim m/\nu$} & \cell{$\sim m/\nu$} & \cell{$\sim m/\nu$} & \cell{?} \\ \hline

    \cell{reciprocity} & \cell{of order 1 for all $m/\nu$, converges to finite limit less than $1/2$ for $m/\nu \to \infty$} & \cell{of order 1 for all $m/\nu$, converges to $1/2$ for $m/\nu \to \infty$} & \cell{decreases with $T$ from a value of order 1 for $T=0$ to that of order $1/R$ for $T=2$} & \cell{?} \\ \hline

    \cell{average clustering coefficient} & \cell{of order 1 for all $m/\nu$, converges to a finite limit for $m/\nu \to \infty$} & \cell{of order 1 for all $m/\nu$, converges to a finite limit for $m/\nu \to \infty$} & \cell{decreases with $T$ from a value of order 1 for $T=0$ to that of order $1/R$ for $T=2$} & \cell{?} \\ \hline
\end{tabular}
\caption{Summary of the behavior of the model studied in this paper in the limit of large network size ($N \sim \nu e^R\to\infty$).}
\label{table:models_characteristics}
\end{table}

We believe that the novel class of networks studied here will be useful as a reference model for scale-free directed networks. Notably, another model of directed hyperbolic networks was proposed recently in \cite{boguna_dir}. There each node attains two  intrinsic popularities (akin to radial coordinates in the hyperbolic disk representation), with respect to in- and out-degree, respectively. By regulating the distributions of these popularities and the correlations between them the authors of \cite{boguna_dir} are able to generate directed networks with arbitrary in- and out-degree distributions and arbitrary correlation between them. In turn, our model is less suitable to fitting arbitrary degree distributions: indeed, out-degree distribution here is fixed. However, it is more naturally geometric, in a sense that it is a natural generalization of well-studied Euclidean $m$-nearest-neighbour graphs\cite{eucl1,eucl2,eucl3,eucl4,balobas,walters}, and it might be a more natural fit to model the cases when one of the two distributions is fixed to be delta-functional or Poisson. One possible way of using these models is to study embedding of directed networks in hyperbolic space. The analogous task for the case of undirected networks is well-developed \cite{embed1,embed2,embed3} but generalization to the directed case is in it infancy (although note the recent paper \cite{kovacs}). Using the discrepancy between embedded directed networks and $m$-NN networks as a loss function characterizing the quality of the embedding seems to be a very exciting perspectives, although it goes beyond the scope of this paper.

\newpage

\section*{Appendix}

In this appendix we provide reference data to be used for the generation of directed networks with given characteristics. In Table \ref{table:reciprocity} we tabulate the dependence of the network reciprocity on temperature, i.e. the master curve to which curves on \fig{TVCR}D converge. Table \ref{table:m_r_stats} summarizes the results of our numerical generation of 27 networks with target values of parameters $N = 10^4, 3\times 10^4, 10^5$, $m=5, 15, 40$ and $r=0.1,0.2,0.4$. Here $R$ and $T$ are the values of the disk radius and temperature calculated from \eqref{nu_est} and from Table \ref{table:reciprocity}, respectively, while $\Delta m/m$ and $\Delta r/r$ are the relative discrepancies between generated and target values of average degree and reciprocity:
\begin{equation}
    \Delta m/m = (m_{gen} - m_{target})/m_{target};\ \ \  \Delta r/r = (r_{gen} - r_{target})/r_{target}
\end{equation}

\begin{table}[ht!]
\centering
\begin{tabular}{||c c||} 
\hline
$T$ & $r$ \\
\hline \hline
	0.0 & 0.50133 \\
	0.1 & 0.49449 \\
	0.2 & 0.48011 \\
	0.3 & 0.45954 \\
	0.4 & 0.43477 \\
	0.5 & 0.40725 \\
	0.6 & 0.37806 \\
	0.7 & 0.34797 \\
	0.8 & 0.31756 \\
	0.9 & 0.28723 \\
	1.0 & 0.25728 \\
	1.1 & 0.22797 \\
	1.2 & 0.19955 \\
	1.3 & 0.17234 \\
	1.4 & 0.14669 \\
	1.5 & 0.12297 \\
	1.6 & 0.10154 \\
	1.7 & 0.08262 \\
	1.8 & 0.06632 \\
	1.9 & 0.05257 \\
	2.0 & 0.04122 \\
\hline
\end{tabular}
\caption{Numerically calculated values of reciprocity $r$ for various values of temperatures $T$ ($N = 50000$).}
\label{table:reciprocity}
\end{table}

\begin{table}[ht!]
\centering
\begin{tabular}{||c c c || c c c c||} 
\hline
$N$ & $m$ & $r$ & $R$ & $T$ & $\Delta m/m$ & $\Delta r/r$\\
\hline \hline
    10000 & 5 & 0.1 & 15.685 & 1.612 & -0.013 & -0.011 \\
	10000 & 5 & 0.2 & 15.685 & 1.198 & -0.012 & 0.020 \\
	10000 & 5 & 0.4 & 15.685 & 0.524 & -0.002 & 0.006 \\
	10000 & 15 & 0.1 & 13.488 & 1.612 & -0.039 & 0.051 \\
	10000 & 15 & 0.2 & 13.488 & 1.198 & 0.019 & -0.013 \\
	10000 & 15 & 0.4 & 13.488 & 0.524 & -0.012 & 0.002 \\
	10000 & 40 & 0.1 & 11.526 & 1.612 & -0.026 & 0.101 \\
	10000 & 40 & 0.2 & 11.526 & 1.198 & -0.022 & 0.020 \\
	10000 & 40 & 0.4 & 11.526 & 0.524 & -0.017 & 0.022 \\
	30000 & 5 & 0.1 & 17.882 & 1.612 & 0.011 & -0.024 \\
	30000 & 5 & 0.2 & 17.882 & 1.198 & 0.003 & 0.002 \\
	30000 & 5 & 0.4 & 17.882 & 0.524 & 0.014 & -0.024 \\
	30000 & 15 & 0.1 & 15.685 & 1.612 & -0.015 & 0.033 \\
	30000 & 15 & 0.2 & 15.685 & 1.198 & -0.015 & 0.008 \\
	30000 & 15 & 0.4 & 15.685 & 0.524 & -0.005 & 0.016 \\
	30000 & 40 & 0.1 & 13.723 & 1.612 & -0.014 & 0.046 \\
	30000 & 40 & 0.2 & 13.723 & 1.198 & 0.010 & -0.002 \\
	30000 & 40 & 0.4 & 13.723 & 0.524 & -0.005 & 0.004 \\
	100000 & 5 & 0.1 & 20.290 & 1.612 & -0.012 & -0.037 \\
	100000 & 5 & 0.2 & 20.290 & 1.198 & -0.003 & 0.007 \\
	100000 & 5 & 0.4 & 20.290 & 0.524 & 0.003 & -0.003 \\
	100000 & 15 & 0.1 & 18.093 & 1.612 & -0.013 & -0.008 \\
	100000 & 15 & 0.2 & 18.093 & 1.198 & 0.010 & -0.013 \\
	100000 & 15 & 0.4 & 18.093 & 0.524 & -0.005 & 0.001 \\
	100000 & 40 & 0.1 & 16.131 & 1.612 & -0.004 & 0.002 \\
	100000 & 40 & 0.2 & 16.131 & 1.198 & -0.010 & 0.009 \\
	100000 & 40 & 0.4 & 16.131 & 0.524 & -0.003 & 0.005 \\
\hline
\end{tabular}
\caption{The results of generating of VCR networks with pre-defined properties. $N$, $m$ and $r$ are the target values of the number of nodes,  average out-(in-)degree and reciprocity, respectively. $R$ and $T$ are the values of disk radius and temperature used for simulations. $\Delta m/m$ and $\Delta r/r$ and relative discrepancies between target values of $m$ and $r$ and their observed values in simulated networks.}
\label{table:m_r_stats}
\end{table}

\newpage

\section*{Acknowledgements}
The authors are grateful to M.\'{A}. Serrano, P. Krapivsky, S. Nechaev, K. Polovnikov and M. Schich for stimulating discussions. This work was partially supported by CUDAN ERA Chair project (Grant No. 810961 of the EU Horizon 2020 
program) and NSF grant No IIS-1741355.


\begin{thebibliography}{99}

\bibitem{viral} Leskovec J., Adamic L.A., Huberman B.A., The dynamics of viral marketing, ACM Transactions on the Web (TWEB), {\bf 1} (1), 5-es (2007).

\bibitem{recom} Oestreicher-Singer G., Sundararajan A.,Recommendation networks and the long tail of electronic commerce, MIS Quaterly, {\bf 36}, 65 (2012).

\bibitem{kiss} Kiss G., Armstrong C., and Milroy R., Piper J., An associative thesaurus of English and its computer analysis. In Aitken A. J. and Bailey R. W., \& Hamilton-Smith N. (Eds.), The Computer and Literary Studies, Edinburgh, UK: Edinburgh University Press, 1973.

\bibitem{nelson} Nelson D.L., McEvoy C.L., and Schreiber T.A., The University of South Florida free association, rhyme, and word fragment norms, Behavior Research Methods, Instruments, \& Computers, {\bf 36}, 402, (2004).

\bibitem{dedayne} De Deyne S., Navarro D.J., Perfors A., Brysbaert M., and Storms G., The “Small World of Words” English word association norms for over 12,000 cue words, Behavior Research Methods, {\bf 51}, 987 (2019).

\bibitem{valba} Valba O., Gorsky A., Nechaev S., Tamm M., ``Analysis of English free association network reveals mechanisms of efficient solution of Remote Association Tests'', PLOS One, {\bf 16} (4): e0248986 (2021).

\bibitem{dedeo} Viteri S., DeDeo S., ``Epistemic phase transitions in mathematical proofs'', Cognition, {\bf 225}, 105120 (2022).

\bibitem{konnect} http://konect.cc/networks, datasets for DBpedia, English wikipedia, zhishi.me. 

\bibitem{wiki} Capocci A.,  Servedio V. D. P., Colaiori F., Buriol L. S., Donato D., Leonardi S., and Caldarelli G., Preferential attachment in the growth of social networks: The internet encyclopedia Wikipedia, Phys. Rev. E {\bf 74}, 036116 (2006).

\bibitem{dorog} Dorogovtsev S.N., Mendes J.F.F., Evolution of Networks: From Biological Nets to the Internet and WWW, Oxford University Press, 2003. 

\bibitem{jackson} Jackson M.O., Social and Economic Networks, Princeton University Press, 2008.

\bibitem{newman_book} Newman M.E.J., Networks, Oxford University Press, 2018.

\bibitem{krapivsky_book} Krapivsky P.L., Redner S., Ben-Naim E., A Kinetic View of Statistical Physics, Cambridge University Press, 2010.

\bibitem{barabasi_book} Barab\'{a}si A.-L., Network Science, Cambridge University Press, 2016.

\bibitem{barthelemy_book} Barthelemy M., Spatial Networks: a Complete Introduction, Springer, 2022.

\bibitem{flory} Flory P.J., Molecular Size Distribution in Three Dimensional Polymers, J. Chem. Phys., {\bf 63}, 3083, 3091, 3096 (1941).

\bibitem{erdos} Erd\H{o}s P., R\'{e}nyi A., On the evolution of random graphs, Publ. Math. Inst. Hung. Acad. Sci, {\bf 5}, 17 (1960).

\bibitem{ws} Watts D.J., Strogatz S.H., Collective dynamics of ‘small-world’ networks, Nature {\bf 393}, 440 (1998).

\bibitem{ba} Barab\'{a}si A.-L., Albert R., Emergence of Scaling in Random Networks, Science, {\bf 286}, 509 (1999). 

\bibitem{krapivsky_pref} Krapivsky P.L., Redner S., Leyvraz F., Connectivity of growing random networks, Phys. Rev. Lett., {\bf 85}, 4629 (2000).

\bibitem{conf} Newman M.E.J., Strogatz S.H, Watts D.J., Random graphs with arbitrary degree distributions and their applications, Phys. Rev. E, {\bf 64}, 026118 (2001).

\bibitem{ParkNewman} Park J., Newman M.E.J., The statistical mechanics of networks, Phys. Rev. E {\bf 70}, 066117
(2004). 

\bibitem{krioukov1} Serrano M.\'{A}., Krioukov D., Bogun\'{a} M., Self-similarity of complex networks and hidden metric spaces, Phys. Rev. Lett. {\bf 100} (7), 078701 (2008).

\bibitem{krioukov2} Krioukov D., Papadopoulos F., Kitsak M., Vahdat A., Bogun\'{a} M., Hyperbolic geometry of complex networks, Phys. Rev. E, {\bf 82} (3), 036106 (2010).

\bibitem{krioukov3} Papadopoulos F., Kitsak M., Serrano M.\'{A}., Bogun\'{a} M., Krioukov D., Popularity versus similarity in growing networks, Nature, {\bf 489}, 537 (2012).

\bibitem{apol1} Andrade J.S., Herrmann H.J., Andrade R.F.S., da Silva, L. Apollonian networks: Simultaneously scale-free, small world,
Euclidean, space filling, and with matching graphs, Phys. Rev. Let., {\bf 94}, 018702 (2005).

\bibitem{apol2} Zhou T., Yan G., Wang B.-H., Maximal planar networks with large clustering coefficient and power-law degree distribution,
Phys. Rev. E, {\bf 71}, 046141 (2005).

\bibitem{zhang1} Zhang Z., Comellas F., Fertin G., Rong L., High-dimensional Apollonian networks, J. Phys. A, {\bf 39}, 1811 (2006).

\bibitem{zhang2} Zhang Z., Rong L., Comellas F., High-dimensional random Apollonian networks, Physica A, 364, 610 (2006).

\bibitem{bianc} Bianconi G., Ziff R.M., Topological percolation on hyperbolic simplicial complexes, Phys. Rev. E, {\bf 98}, 052308 (2018).

\bibitem{tks} Tamm M.V., Koval D.G., Stadnichuk V.I., Polygon-based hierarchical planar networks based on generalized Apollonian construction, MDPI Physics, {\bf 3}, 998 (2021).

\bibitem{boguna_dir} Allard A., Serrano M.\'{A}., Bogun\'{a} M., Geometric description of clustering in directed networks, arxiv:2302.09055 (2023).

\bibitem{kovacs} Kov\'{a}cs B., Palla G., Model-independent embedding of directed networks into Euclidean and hyperbolic spaces, Communications Physics, {\bf 6}, 28 (2023).

\bibitem{eucl1}  Miller G.L., Teng S.H., Vavasis S.A., An unified geometric approach to graph separators, in IEEE 32nd Annual Symposium on Foundations of Computer Science, 538 (1991).

\bibitem{eucl2} H\"{a}ggstr\"{o}m  O., Meester R., Nearest neighbor and hard sphere models in continuum percolation, Random Structures and Algorithms, {\bf  9}, 295 (1996).

\bibitem{eucl3} Balister P., Bollob\'{a}s B., Sarkar A., Walters M., Connectivity of random k-nearest neighbour graphs, Advances in Applied Probability, {\bf 37}, 1  (2005).

\bibitem{eucl4}  Teng S., Yao F., k-nearest-neighbor clustering and percolation theory, Algorithmica, {\bf 49}, 192 (2007).	

\bibitem{balobas} Balister R., Sarkar A., Balob\'{a}s B., Handbook of Large Scale Random Networks, 2008.

\bibitem{walters} Walters M., Random geometric graphs, p. 365 in "Surveys in Combinatorics", ed. R. Chapman, London Mathematical Society Lecture Notes Series, v. 392, 2011.

\bibitem{diff} Coifman R.R., Lafon S.A., Lee B., Maggioni M., Nadler B., Warner F., Zucker F.B., Geometric diffusions as a tool for harmonic analysis and structure definition of data: diffusion maps.
Proc. Natl. Acad. Sci., {\bf  102}, 7426–7431 (2005); Coifman R.R., Lafon S., Diffusion maps. Appl. Comput. Harmon. Anal., {\bf 21}, 5–30
(2006).

\bibitem{tsne} Van der Maaten L., Hinton G., Visualizing data using t-SNE, Journal of Machine Learning Research, {\bf 9}, 2579 (2008)

\bibitem{umap} McInnes L., Healy J., Melville J., UMAP: Uniform Manifold Approximation and Projection for Dimension Reduction, arXiv:1802.03426 (2018).

\bibitem{cannon} Cannon J.W., Floyd W.J., Kenyon R., Parry W.R., Hyperbolic geometry, in Flavors of geometry, ed. by S. Levy, Cambridge University Press, 1997.

\bibitem{taowu} Tao R., Wu F.Y., The vicious neighbour problem, J. Phys. A, {\bf 20}, L299 (1987).

\bibitem{gracar} Gracar P., Grauer A, L\"{u}chtrath L., M\"{o}rters P., The age-dependent random connection model, Queing Systems, {\bf 93}, 309 (2019).

\bibitem{Garlaschelli} Garlaschelli D., Loffredo M.I., "Patterns of Link Reciprocity in Directed Networks", Phys. Rev. Letters, {\bf 93}, 268701 (2004).

\bibitem{alon1} Milo R., Shen-Orr S., Itzkovitz S., Kashtan N., Chklovskii D., Alon U., "Network motifs: simple building blocks of complex networks", Science, 298, 824 (2002).

\bibitem{alon2} Milo R., Itzkovitz S., Kashtan N., Levitt R., Shen-Orr S. et al., "Superfamilies of evolved and designed networks", Science, 303, 1538 (2004).

\bibitem{Newman_recip} Newman M.E.J., Forrest S., Balthrop J., "Email networks and the spread of computer viruses", Phys. Rev. E, {\bf 66}, 035101(R) (2002).

\bibitem{citations}  De Solla Price D.J., Networks of Scientific Papers, Science, {\bf 149}, 510 (1965).

\bibitem{lgk} Lifshitz I.M., Grosberg A.Y., Khokhlov A.R., Some problems of the statistical physics of polymer chains with volume interaction, Rev. Mod. Phys. {\bf 50}, 683 (1978).
 
\bibitem{erukh1} Erukhimovich I., Thamm M.V., Ermoshkin A.V., Theory of the sol-gel transition in thermoreversible gels with due regard for the fundamental role of mesoscopic cyclization effects, Macromolecules, 34, 5653 (2001).


\bibitem{embed1} Nickel M., Kiela D., Poincar\'{e} Embeddings for Learning Hierarchical Representations, Proceedings of Advances in Neural Information Processing Systems 30 (NIPS 2017), arXiv:1705.08039.

\bibitem{embed2} Sala F., De Sa C., Gu A., R\'{e} C., Representation Tradeoffs for Hyperbolic Embeddings, Proceedings of the 35th International Conference on Machine Learning, PMLR {\bf 80}, 4460 (2018).

\bibitem{embed3} Tifrea A., B\'{e}cigneul G., Ganea O.-E., Poincar\'{e} GloVe: Hyperbolic Word Embeddings, Proceedings of the International Conference on Learning Representations (ICLR 2019), arXiv:1810.06546.

\end{thebibliography}
\end{document}